\documentclass[conference,10pt]{IEEEtran}
\IEEEoverridecommandlockouts
\usepackage{cite}
\usepackage{amsmath,amssymb,amsfonts}
\usepackage{algorithmic}
\usepackage{graphicx}
\usepackage{textcomp}
\usepackage{braket}
\def\BibTeX{{\rm B\kern-.05em{\sc i\kern-.025em b}\kern-.08em
    T\kern-.1667em\lower.7ex\hbox{E}\kern-.125emX}}

\UseRawInputEncoding

\begin{document}

\title{Infomechanics of Independent and Identically Distributed Particles
\\
}

\author{Arnaldo Spalvieri$^*$ \\
$^*$Dipartimento di Elettronica, Informazione e Bioingegneria,
Politecnico di Milano
}

\maketitle

\begin{abstract}
The paper moves a step towards the full integration of statistical
mechanics and information theory. Starting from the assumption
that the thermodynamical system is composed by particles whose
quantized energies can be modelled as independent and identically
distributed random variables, the paper proposes an approach whose
cornerstones are the information-theoretic typical set and the
conditional equiprobability of microstates given certain
macrostates of the system. When taken together, these two concepts
explain why the standard assumption of equally probable
microstates is non-necessary, if not misleading. Several new
specific results of physical relevance are derived from this
approach, among which are the probability distribution of the
occupancy numbers of the quantum states and an exact formula for
the ideal gas in a container that gives the entropy of the gas
also at low temperature. These specific results are pieces of a
self-consistent and unified framework that encompasses the cases
of low and high temperature, of distinguishable and
indistinguishable particles, of small and large number of
particles.
\end{abstract}

\section{Introduction}

The following passage of an interview with Shannon can be found
in \cite{tribus}: {\em My greatest concern was what to call it. I
thought of calling it 'information,' but the word was overly used,
so I decided to call it 'uncertainty.' When I discussed it with
John von Neumann, he had a better idea. Von Neumann told me, ''You
should call it entropy, for two reasons. In the first place your
uncertainty function has been used in statistical mechanics under
that name, so it already has a name. In the second place, and more
important, no one really knows what entropy really is, so in a
debate you will always have the advantage.''}

Starting from the concept of entropy that Information Theory (IT)
and Statistical Mechanics (SM) share, many authors built in the
past bridges between them. Among the works that more contributed
to the strengthen link between the two disciplines we cite the
paper of Landauer \cite{landauer}, that showed the equivalence
between heat and logical information. This major result is today
proved experimentally \cite{landdemonstration} and it is widely
accepted that (quantum) thermodynamics can be treated by (quantum)
information-theoretic tools, see for instance the tutorial paper
\cite{plenio}. With reference to the connections between classical
thermodynamics and classical information theory, we limit
ourselves to mention the recent survey \cite{mariosurvey}, the
pioneering work of Jaynes \cite{maxent} and the large attention
that has been attracted in the past by the connections between
information theoretic inequalities and the irreversibility of
certain thermodynamical processes, see \cite{merhav} and Chapter 4
of \cite{cover}. Recently, the development of quantum information
theory has brought the community of information theory researchers
closer to quantum thermodynamics than ever before, see for
instance the recent tutorial \cite{itthermoreview}. Making a
comprehensive review of the bibliography that links information
theory and thermodynamics and, more generally, of the extremely
vast bibliography that, in various manners, touches the topics
that are teated in this paper, is out of our scope. We will limit
ourselves to cite and to comment along the paper the bibliography
that we find relevant to the specific point at hand.

Despite the evidence that heath and logical information are
equivalent, still today the standard textbook approach to
statistical mechanics and thermal physics is strongly physically
oriented, leaving to classical and quantum information theory only
a minor role. Although many textbooks have a chapter or a section
devoted to information theory, quantities as random entropy,
conditional entropy, mutual information, Holevo accessible
information, are often overlooked. This happens because these
quantities and the associated concepts, that are of fundamental
importance in information theory, actually seem not to have yet
found their role in classical and quantum statistical mechanics.
The identification of these information-theoretic quantities with
the corresponding physical quantities is the vision that guides
the research reported in this paper. In our understanding, more
than proposing a theory, this paper proposes a vision, the
infomechanical vision, where the hybridization between the two
disciplines is deep as never it has been before.

Assuming that particles can be modelled as independent subsystems
and that the probability distribution of particle's quantum state
is the same for all the particles, and starting from the guiding
idea that macrostates are random global attributes of the system
that play the role of random conditions in the conditional
probability of microstates, we {\em demonstrate} the following
specific results of physical relevance that, to our best
knowledge, are new:
\begin{itemize}
\item  the probability distribution of the occupancy numbers in a
system of a fixed number of particles that can take many energy
values is the multinomial distribution, \item  microstates of the
above system are conditionally equiprobable given the occupancy
numbers,  \item an exact formula for the entropy of the ideal gas
in a container, that holds with equality also at low temperature.
\end{itemize}
These specific results are pieces of a puzzle where low and high
temperature, indexed (distinguishable) and non-indexed
(indistinguishable) particles, small and large number of
particles, coexist and find their own place in harmony between
them.

The outline of the paper is as follows. In section II we present
the notation and the main definitions. Although this material is
standard in information theory, we present it because, sice it
could be not part of the background of researchers of statistical
mechanics, it contributes to make the paper self-contained.
Section III presents the assumptions that we make and the system
model. Here we pass from the representation of the system in the
phase space to the representation of the system in the space of
its quantum states. This leads to a natively quantum definition of
entropy, that does not requires quantization of the phase space
and the subsequent need for approximations at low temperature, see
e.g. \cite{psapprox}. In section IV, systems made by non-indexed
particles are considered. The main conceptual point of this
section is the identification of the entropy of indistinguishable
particles with the entropy of the occupancy macrostate. The
crucial point from the technical point of view is the observation
that the probability distribution of the occupancy numbers is the
multinomial distribution, a result that, as claimed before, seems
to be new. Section V is devoted to the application of the findings
of section IV to systems at the thermal equilibrium.  One specific
instance of non-indexed particles is the ideal dilute gas in a
container, that is studied in section VI, where it is given an
exact entropy formula that describes the thermodynamics of the gas
both at low and at high temperature. In section VII the main
points of the paper are summarized and comments about open points
that could be addressed in the future are presented. The
appendixes address specific points and propose historical remarks
that can be skipped at the first reading.

\section{Notation and definitions}

Let the uppercase calligraphic character, e.g. ${\cal L}$, denote
a random variable and let $\{ {\cal L} \}$ be the support set of
the random variable. Unless explicitly stated, in the following we
will refer to discrete random variables. We denote $|{\cal L}|$
the number of elements of the support set of the discrete random
variable and we indicate the probability $Pr({\cal L}=l)$ with the
shorthand $p_{\cal L}(l)$, that is
\[p_{\cal L}(l) \stackrel{\text{def}}{=} Pr({\cal L}=l), \ \
\forall \ l \in \{{\cal L}\}.\] The probability distribution, or,
simply, distribution, of the random ${\cal L}$, is the set
$\{p_{\cal L}(l)\}$, $\forall \ l \in \{{\cal L}\}$, that will be
often hereafter indicated with the shorthand $p_{\cal L}$. In the
following we will use the angular brackets to denote the
expectation (when it exists) of a function of random variable with
respect to the probability distribution of the random variable
(the calligraphic character) that is inside the brackets:
\[\braket{\epsilon({\cal L})} \stackrel{\text{def}}{=}\sum_{l \in \{{\cal L}\}}p_{\cal
L}(l)\epsilon(l),\] where $\epsilon(\cdot)$ is a deterministic
function of its argument. The above notation can be extended in a
straightforward way to the case of functions of many random
variables, by taking the expectation over the joint probability
distribution of the many random variables that are inside the
angular brackets. Let us consider the random variable
\[{\cal E}=\epsilon({\cal L}).\]
When the context leaves no room to ambiguity about the
deterministic function $\epsilon(\cdot)$ that maps ${\cal L}$ onto
${\cal E}$, we will use the familiar notation $\mu_{\cal E}$ and
$\sigma^2_{\cal E}$ for mean value and variance of ${\cal E}$:
\begin{equation}\mu_{{\cal E}} \stackrel{\text{def}}{=} \braket{\epsilon({\cal L})},
\nonumber\end{equation}
\[\sigma^2_{{\cal E}} \stackrel{\text{def}}{=} \braket{\epsilon^2({\cal L})}-(\braket{\epsilon({\cal L})})^2.\]
In general, when we want to shorten the notation of expectation we
put the random variable in the subscript, while when we want to
indicate a deterministic function of a random variable we put the
random variable inside the round bracket. The special case
$\epsilon(\cdot)=\delta(\cdot)$, where $\delta(\cdot)$ is the
indicator function,
\begin{align} & \hspace{-0.2cm} \delta(x)=\left\{
\begin{array}{cc} 1, & \ \ x=0, \\
0, & \ \ x \neq 0,\\
\end{array} \right.
\nonumber\end{align} shows two interpretations of the function
$p_{\cal L}(\cdot)$. If the argument of $p_{\cal L}(\cdot)$ is
deterministic, then the deterministic $p_{\cal L}(\cdot)$
expresses probability as expectation:
\begin{align}
p_{\cal L}(l)&= \sum_{k \in \{{\cal L}\}}p_{\cal L}(k)\delta(k-l)
\label{detp}
\\ &=\braket{\delta({\cal L}-l)}. \nonumber
\end{align}
Instead, if the argument of $p_{\cal L}(\cdot)$ is random, then
$p_{\cal L}(\cdot)$ is the random variable that is obtained by
mapping the outcome of the random argument into its probability:
\begin{align}
p_{\cal L}({\cal L})= \sum_{l \in \{{\cal L}\}}p_{\cal L}(l)
\delta(l-{\cal L}). \label{ranp}
\end{align}
The reader could find the above discussion about the meaning of
$p_{\cal L}(l)$ and $p_{\cal L}({\cal L})$ redundant, if not
trivial, but we want to stress the difference between $p_{\cal
L}(l)$ and $p_{\cal L}({\cal L})$ because, in this paper, this
difference is crucial: in the rest of the paper, $p_{\cal L}(l)$
means the deterministic quantity (\ref{detp}) while $p_{\cal
L}({\cal L})$ means the random quantity (\ref{ranp}).

\subsection*{Surprise, uncertainty, information}

In classical information and communication theory, the {\em
random} entropy $H({\cal L})$ is the following deterministic
function of the discrete random variable ${\cal L}$:
\begin{equation}
\infty > H({\cal L}) \stackrel{\text{def}}{=} -k \log(p_{\cal
L}({\cal L})) \geq 0,\nonumber
\end{equation}
where $k>0$ is a constant that depends on the context and the
inequalities are obvious because $0 < p_{\cal L}({\cal L}) \leq 1$
(we exclude from the support set of ${\cal L}$ events with zero
probability to guarantee that the random entropy is limited). In
physics $k=k_B$, while, in information and communication theory,
\[k=\frac{1}{\log(2)},\]
or, equivalently, one takes the base-2 logarithm and $k=1$. In
what follows,  we drop the multiplicative constant when it is
unnecessary, being understood that it must be recovered from the
context when it becomes necessary. Information and communication
theory researchers often call $H({\cal L})$ {\em surprise},
because it quantitatively expresses the surprise that one
experiences when he observes the result ${\cal L}$ of the random
experiment. Being a deterministic function of a random variable,
see (\ref{ranp}) and the relevant discussion about its meaning,
the above $H({\cal L})$ is a random variable itself. The
expectation of $H({\cal L})$ with respect to the probability
distribution $p_{\cal L}$ is the Shannon entropy of ${\cal L}$:
\begin{equation} H_{\cal L}
\stackrel{\text{def}}{=} -\sum_{l \in \{{\cal L}\}}p_{\cal
L}(l)\log(p_{\cal L}(l)) \geq 0,\nonumber\end{equation} where we
use the shorthand notation
\[H_{\cal L}=\braket{H({\cal L})}\]
and the inequality becomes equality only when ${\cal L}$ is
non-random.

The extension to random vectors is straightforward. The random
entropy of the random vector $\bar{\cal L}$ is
\begin{equation}H(\bar{\cal L}) \stackrel{\text{def}}{=} -\log(p_{\bar{\cal L}}(\bar{l})).\label{systemsurprise}
\end{equation}
Its expectation, that is the Shannon entropy of the random vector,
is
\begin{align} H_{\bar{\cal L}}& \stackrel{\text{def}}{=}
-\sum_{\bar{l} \in \{\bar{\cal L}\}} p_{\bar{\cal L}}(\bar{l})
\log(p_{\bar{\cal L}}(\bar{l}))\label{gibbs}.
\end{align}
 The above entropy is called in
thermodynamics {\em Gibbs entropy}.

Another approach to entropy in thermodynamics is the following.
Let the $l$-th random occupancy number ${\cal N}_{\Sigma}(l)$ be
the number of elements of $\bar{\cal L}^N$ whose value is $l$:
\begin{equation}{\cal N}_{\Sigma}(l) \stackrel{\text{def}}{=} \sum_{i=1}^N \delta(l
-{\cal L}_i),  \ \ \forall \ l \ \in \{{\cal L}\},\label{on}
\end{equation}
with
\begin{equation}
\sum_{l \in \{{\cal L}\}}{\cal N}_{\Sigma}(l)=N.
\nonumber\end{equation} Assuming that the entries of the random
vector $\bar{{\cal L}^N}$ are independent and identically
distributed (i.i.d) discrete random variables, one can write
\begin{align}H_{\bar{\cal L}^N}&=-N\sum_lp_{\cal L}(l)\log(p_{\cal L}(l))
\nonumber \\ &= -\sum_l \langle{\cal N}_{\Sigma}(l)\rangle \log
\left(\frac{\langle{\cal
N}_{\Sigma}(l)\rangle}{N} \right) \nonumber \\
&\stackrel{\text{Stirling}}{\approx} \log
\left(\frac{N!}{\prod_l\langle{\cal
N}_{\Sigma}(l)\rangle!}\right)=\log \left(W_M(\langle \bar{\cal N}
_{\Sigma}\rangle)\right), \label{shmulti}
\end{align}
where $W_M(\cdot)$ is the nultinomial coefficient,
\[\langle{\cal
N}_{\Sigma}(l)\rangle=Np_{\cal L}(l) \] is the expectation of the
$l$-th random occupancy number and
\[\log(n!)\stackrel{\text{Stirling}}{\approx}n(\log(n)-1)\] is Stirling's
approximation. The gamma function can be used in place of the
factorial when $\langle{\cal N}_{\Sigma}(l)\rangle$ is
non-integer. The multinomial coefficient is interpreted in
standard combinatorics as the number of distinct permutations of
the elements of the vector, where distinct means that if the
positions of two elements with the same value are permuted, then
the vector before and after the permutation is the same and the
permutation must not be counted as a distinct permutation. For
instance, the number of distinct permutations of a vector of four
elements with two zeros and two ones is six. When all the elements
have distinct values, as it happens with unit probability when the
$N$ elements are the outcomes of continuous random variables, all
the permutations are distinct and the multinomial coefficient is
equal to $N!$. Although the identification of the above
multinomial coefficient with the $W$ appearing in (\ref{bp}) leads
to the identification of the Boltzmann-Planck entropy with the
Gibbs entropy, see e.g. eqn. (18.14) of \cite{sek}, still today
researchers are actively trying to put in a coherent relation the
two, see e.g. \cite{gbagain}. The better succeeded of these
attempts is, to our opinion, paper \cite{zupa}. We will return on
\cite{zupa} in Appendix B to discuss the relation between this
paper and \cite{zupa}.

Given two random variables ${\cal L}$ and ${\cal X}$, the
conditional random entropy of ${\cal L}$ given ${\cal X}$ is
\[H({\cal L}|{\cal X}) \stackrel{\text{def}}{=} -
\log(p_{{\cal L}|{\cal X}}({\cal L},{\cal X})),\] where the
familiar notation is used for the conditional probability
distribution. The expectation of the conditional random entropy
over the joint probability distribution of ${\cal L}$ and ${\cal
X}$ is the conditional entropy of ${\cal L}$ given ${\cal X}$,
that is
\begin{equation}H_{{\cal L}} \geq H_{{\cal L}|{\cal X}}
\stackrel{\text{def}}{=} \sum_{y \in \{{\cal Y}\}} p_{\cal
X}(x)H_{{\cal L}|x} \geq 0,\label{centrdef}
\end{equation} where
\[H_{{\cal L}|x} \stackrel{\text{def}}{=}
-\sum_{l \in \{{\cal L}\}} p_{{\cal L}|{\cal X}}(l,x)
\log(p_{{\cal L}|{\cal X}}(l,x))\] is the conditional entropy of
${\cal L}$ with deterministic condition ${\cal X}=x$, see chapter
2 of \cite{cover} for the definition of conditional entropy and
for the inequalities in (\ref{centrdef}). The conditional entropy
of ${\cal L}$ given ${\cal X}$ is a measure of the uncertainty
that we have about the unknown ${\cal L}$ given the outcome of
${\cal X}$. One of the most fundamental quantities in classical
information theory is the {\em information} $I_{{\cal L};{\cal
X}}$ between ${\cal L}$ and ${\cal X}$:
\begin{equation}I_{{\cal L};{\cal X}} \stackrel{\text{def}}{=}
H_{{\cal L}}-H_{{\cal L}|{\cal X}}\geq 0, \nonumber \end{equation}
where the inequality becomes equality only when ${\cal L}$ and
${\cal X}$ are independent random variables, in which case the
uncertainty that we have about the unknown ${\cal L}$ is unchanged
by the given ${\cal X}$. The above quantity, introduced by Shannon
in \cite{shannon}, is interpreted as the reduction of uncertainty
about ${\cal L}$ achieved thanks to the given ${\cal X}$, that is,
the information that ${\cal X}$ brings about the unknown ${\cal
L}$. By straightforward use of Bayes rule one has
\begin{equation}I_{{\cal L};{\cal X}}=H_{{\cal L}}-H_{{\cal L}|{\cal X}}=H_{{\cal
X}}-H_{{\cal X}|{\cal L}}=I_{{\cal X};{\cal L}}.
\label{mutual}\end{equation} For this reason, often the
information is called {\em mutual information}. The extension to
random vectors is straightforward and is not pursued for brevity.

\section{System model}

One quantum particle is modelled as the probabilistic mixture of
pure states and is represented by the density operator
\begin{equation}\hat{\rho}=\sum_{l \in \{{\cal L}\}} p_{{\cal
L}}({l})\ket{\phi({l})}\bra{\phi({l})}, \label{rho1}
\end{equation} where the bra-ket notation is used for quantum
states and
\[\{\ket{\phi({l})}\},
\ \ {{l}} \in \{{{\cal L}}\} \] are the energy eigenstates
resulting from the solution of the one-particle time-independent
Schr\"{o}dinger equation. Since the eigenstates
$\{\ket{\phi({l})}\}$ forms an orthonormal basis of the Hilbert
space of the one-particle system, equation (\ref{rho1}) gives the
density operator in the form of its spectral decomposition. In
this case, the Von Neumann entropy $S(\hat{\rho})$ coincides with
the Shannon entropy of the classical random variable ${\cal L}$:
\begin{equation}S(\hat{\rho})\stackrel{\text{def}}{=}
- \mbox{Tr} (\hat{\rho}\log(\hat{\rho}))=H_{\cal L}.\label{vne}
\end{equation}
In this model, the particle takes discrete energy values that
belong to the set $\{\epsilon(l)\}$ of the eigenvalues associated
to the eigenstates. One popular example is the set of energy
eigenvalues of the quantum harmonic oscillator, where
\begin{equation}\epsilon(l)
= l \omega \hbar+\frac{1}{2}\omega \hbar, \ \  l \ \in
\{0,1,\cdots\},\label{pib0}
\end{equation} where $l$ is the quantum number,
\[\hbar=1.05 \cdot 10^{-34} \ \ \mbox{J} \cdot \mbox{s}\]
is the reduced Planck constant, $\omega$ is the angular frequency
of the oscillation and
\[\epsilon(0)=\frac{1}{2}\omega \hbar\]
is the energy of the ground state.
Another example is that of a particle of mass $m$ in an
one-dimensional box of length $L$, for which the solutions of the
time-independent Schr\"{o}dinger equation with aperiodic boundary
conditions have eigenvalues
\begin{equation}\epsilon(l)=
l^2 \frac{\pi^2 \hbar^2}{2 m L^2},\ \ l \ \in
\{1,2,\cdots\}.\label{pib}
\end{equation}

A system made by $N$ quantum particles of the same species is
modelled by the density operator
\begin{align}\hat{\rho} &=\sum_{\bar{l} \in \{\bar{\cal L}\}}p_{\bar{\cal
L}}(\bar{l})\ket{\phi(\bar{l})}\bra{\phi(\bar{l})},
\label{mixedsystem2}
\end{align}
where  $\{\ket{\phi(\bar{l})}\}$ form an orthonormal basis of the
Hilbert space of the system, therefore
\begin{equation}S(\hat{\rho})=H_{\bar{\cal L}^N}.
\label{vne3}
\end{equation}
 Vector $\bar{l}$ completely
describes the system at the microscopic level. For this reason, in
statistical mechanics and thermal physics, $\bar{l}$ is called the
{\em micro}state of the system, see for instance 1.1. of
\cite{pathria}. Distinct microstates of the system are obtained by
permuting any two elements of the vector, unless the values of the
two permuted elements are the same, in which case the permutation
does not lead to a distinct microstate.

Our main assumption is that the $N$ classical random variables of
the random vector $\bar{\cal L}$ are independent and identically
distributed (i.i.d.):
\begin{equation}
p_{\bar{\cal{L}}}(\bar{l})=\prod_{i=1}^Np_{{\cal{L}}_i}(l_i),
\label{indep}
\end{equation}
\begin{equation}p_{{\cal{L}}_1}(l)=p_{{\cal{L}}_2}(l)= \cdots =
p_{{\cal{L}}_N}(l)=p_{{\cal{L}}}(l).\nonumber 
\end{equation}
Writing the basis states of system's Hilbert space as the tensor
product of the basis states of the Hilbert subspaces of the
individual particles, we see that the density operator
$\hat{\rho}$ can be factored into the density operators of the $N$
particles:
\begin{align}\hat{\rho}&=  \hat{\rho}_1 \otimes \hat{\rho}_2 \otimes \cdots \otimes
\hat{\rho}_N,\label{seppart}
\end{align}
where $\otimes$ denotes the tensor product,
\[\hat{\rho}_i=\sum_{l \in \{{\cal L}\}} p_{{\cal
L}}({l})\ket{\phi({l})}_{i}\bra{\phi({l})}_i, \ \ i=1,2,
\cdots,N,\] is the density operator of the i-th particle and
\[\{\ket{\phi({l})}_i\},
\ \ {{l}} \in \{{{\cal L}}\}, \ \ i=1,2,\cdots,N,\] form an
orthonormal basis for the Hilbert subspace of the $i$-th particle.
Equation (\ref{seppart}), which means that particles of the
ensemble represented by $\hat{\rho}$ are non entangled between
them, imports inside $\hat{\rho}$ the standard assumption of
non-interacting particles. Identical distribution seems to be a
physically sound assumption for systems at the thermal equilibrium
even if, in principle, this assumption is an abstract one that
could not limit the system model to systems at the thermal
equilibrium. From (\ref{vne3}) and from the i.i.d. assumption it
follows that
\begin{equation}S(\hat{\rho})=NH_{{\cal L}}.
\label{vne2}
\end{equation}
Other very well known consequences of the i.i.d. assumption are
that the total energy of the system is the sum of the energies of
the particles, the partition function of the system is the product
of the partition functions of the individual particles, the
distribution of system's energy that maximizes the entropy of the
system is the Boltzmann distribution.

\section{Macrostates}

In this section we consider a system made by a deterministic
number $N$ of particles, the case of a random number of particles
being treated in Appendix A. We assume that the size $|{\cal L}|$
of the set of quantum state numbers is finite\footnote{When the
number of quantum state numbers is infinite, to obtain a vector of
occupancy numbers with a finite number of elements we truncate the
set $\{{\cal L}\}$ to a subset $\{{\cal L}'\}$ with finite number
of elements. The truncation is such that
\[Pr({\cal L} \not\in \{{\cal L}'\})  \]
is small enough for our purposes. When the event ${\cal L} \not\in
\{{\cal L}'\}$ happens, we force ${\cal L}$ to the closest energy
level in $\{{\cal L}'\}$. This transformation of ${\cal L}$
produces a new random vector ${\cal L}'$ that we consider in place
of ${\cal L}$. The probability distribution of ${\cal L}'$ and the
occupancy numbers follow in a straightforward way.} and call
$\bar{n}_{\Sigma}$  the generic vector of the $|{\cal L}|$
occupancy numbers, see (\ref{on}) for the definition of the
occupancy number. For any given vector $\bar{n}_{\Sigma}$ of
occupancy numbers, the {\em occupancy macrostate} is the set of
microstates
\begin{align}
\{\bar{l}(\bar{n}_{\Sigma})\}=\{\bar{l}_1(\bar{n}_{\Sigma}),
 \bar{l}_2(\bar{n}_{\Sigma}),
 \cdots, \bar{l}_{W_M(\bar{n}_{\Sigma})}(\bar{n}_{\Sigma})\},
 \label{permset}
\end{align} where
$\bar{l}_1(\bar{n}_{\Sigma})$ is a vector whose occupancy numbers
are the elements of $\bar{n}_{\Sigma}$, the subscript $_i$
indicates the $i$-th distinct permutation of the $N$ elements of
$\bar{l}_1(\bar{n}_{\Sigma})$,\footnote{As in the case of
microstates, also here the permutation of two particles with the
same quantum number is not a distinct permutation.} and the number
of elements $W_M(\bar{n}_{\Sigma})$ of the set of distinct
permutations is the multinomial coefficient, which has been
already introduced in (\ref{shmulti}).

In the case of non-indexed particles, often referred to as {\em
indistinguishable particles}, the permutation of any two particles
doesn't change the state of the system. In this case, the system
is modelled by the random vector $\bar{\cal N}_{\Sigma}$ of the
occupancy numbers and system's physical entropy is $H_{\bar{\cal
N}_{\Sigma}}$ multiplied by Boltzmann's constant.

\subsection*{Multinomial distribution of the occupancy numbers}

A consequence of the i.i.d. assumption that we make for the random
energy levels is that the probability distribution of the random
vector $\bar{\cal N}_{\Sigma}$ is the multinomial distribution:
\begin{align}  p_{\bar{\cal N}_{\Sigma}}(\bar{n}_{\Sigma})&=
W_M(\bar{n}_{\Sigma})\prod_{l \in \{{\cal L}\}}(p_{\cal
L}(l))^{{n}_{\Sigma}(l)} \label{md} .\end{align} Randomness of
macrostates is discussed in Appendix B.  To our best knowledge, it
is observed here for the first time that the distribution of the
occupancy numbers is the multinomial distribution (\ref{md}).
Although the multinomial coefficient is a recurring figure in
statistical mechanics since the times of Boltzmann, the term
$\prod_{l \in \{{\cal L}\}}(p_{\cal L}(l))^{n_{\Sigma}(l)}$ that
appears in (\ref{md}) is not present in Boltzmann neither, to our
best knowledge, in the subsequent literature, at least with the
meaning that it has in (\ref{md}). It could seem obvious, but it
is worth remarking that, while the multinomial coefficient is a
number, the product between the multinomial coefficient and the
probability distribution $\{\prod_{l \in \{{\cal L}\}}(p_{\cal
L}(l))^{n_{\Sigma}(l)}\}$ is a probability distribution!  In the
open literature, the multinomial distribution is considered only
when the probability distribution $p_{\cal L}$ is uniform. In this
case, see e.g. \cite{niven,safra} and II.5 of \cite{feller}, the
uniform distribution of course is not the distribution of energy
levels. The multinomial distribution appears in eqn. (8) of 6.2 of
\cite{pathria} with the same meaning as here, but in the cited
reference it appears only in a passage of a mere algebraic
manipulation of a partition function that allows to calculate the
partition function itself by the help of the multinomial theorem.
Actually, the authors of \cite{pathria} seem not to give to the
multinomial distribution any other meaning than that of a term
involved in an algebraic manipulation. In \cite{pathria}, eqn.
6.3.12, the authors claim that the distribution of the occupancy
numbers in the canonical ensemble is the Poisson distribution.
However, the occupancy numbers cannot follow the Poisson
distribution if the total number of particles is deterministic, as
it actually is in the canonical ensemble, hence this claim is not
compatible with the constraint on the total number of particle of
eqn. 6.2.3 of the same book. An interesting use of the multinomial
distribution is made in \cite{swe}, but, in that paper, entropy is
defined in the phase space and the multinomial distribution is
used to describe the occupancy of space, not of the energy levels.

\subsection*{Conditional equiprobability of microstates given the
occupancy numbers}

Since
\[\prod_{l \in \{{\cal L}\}}(p_{\cal L}(l))^{{n}_{\Sigma}(l)}
=p_{\bar{\cal L}}(\bar{l}), \ \ \ \ \forall \ \bar{l} \in
\{\bar{l}(\bar{n}_{\Sigma})\},\]  equation (\ref{md}) can be
written as
\begin{align}  p_{\bar{\cal L}}(\bar{l})&=
 (W_M(\bar{n}_{\Sigma}))^{-1} p_{\bar{\cal N}_{\Sigma}}(\bar{n}_{\Sigma}), \ \ \ \ \forall \ \bar{l} \in
\{\bar{l}(\bar{n}_{\Sigma})\},\label{t1}\end{align} which shows
that microstates' probability depends only on the occupancy
numbers, or, in other words, that microstates are equiprobable
given the occupancy numbers. We express this concept by saying
that microstates are {\em conditionally equiprobable} given the
occupancy numbers.

To obtain a formal expression of conditional equiprobability we
must use conditional probability distributions. Since $\bar{\cal
N}_{\Sigma}$ is known given $\bar{\cal L}$, the conditional
probability distribution of $\bar{\cal N}_{\Sigma}$ given
$\bar{\cal L}$ is
\begin{align} \hspace{-0.2cm} p_{\bar{\cal N}_{\Sigma}|\bar{\cal L}}(\bar{n}_{\Sigma},\bar{l})=\left\{
\begin{array}{cc} 1, &   \bar{l} \in \{\bar{l}(\bar{n}_{\Sigma})\}, \\
0,  &  \bar{l} \not\in \{\bar{l}(\bar{n}_{\Sigma})\},\\
\end{array} \right.
\label{multcond}\end{align} hence, by Bayes formula,
\begin{align}  p_{\bar{\cal L}|\bar{\cal N}_{\Sigma}}(\bar{l},\bar{n}_{\Sigma})=\left\{
\begin{array}{cc} \frac{p_{\bar{\cal L}}(\bar{l})}
{p_{\bar{\cal N}_{\Sigma}}(\bar{n}_{\Sigma})}, &   \bar{l} \in \{\bar{l}(\bar{n}_{\Sigma})\}, \\
0,  &  \bar{l} \not\in \{\bar{l}(\bar{n}_{\Sigma})\}.\\
\end{array} \right.
\label{bayes}\end{align} Substituting (\ref{t1}) in (\ref{bayes})
we get
\begin{align}  p_{\bar{\cal L}|\bar{\cal N}_{\Sigma}}(\bar{l},\bar{n}_{\Sigma})=\left\{
\begin{array}{cc} \frac{1}{W_M(\bar{n}_{\Sigma})}, &   \bar{l} \in \{\bar{l}(\bar{n}_{\Sigma})\}, \\
0,  &  \bar{l} \not\in \{\bar{l}(\bar{n}_{\Sigma})\},\\
\end{array} \right.
\label{deltaw}\end{align} which shows that microstates are
conditionally equiprobable given the occupancy numbers. Note that
conditional equiprobability is here {\em demonstrated} starting
from the i.i.d. assumption of the elements of $\bar{\cal L}$.

\subsection*{Entropy, conditional entropy and mutual information}

For the conditional random entropies associated to the conditional
probabilities (\ref{multcond}) and (\ref{deltaw}) we have
\begin{align} H(\bar{\cal N}_{\Sigma}|\bar{\cal L})=
0, \nonumber
\end{align}
\begin{align} H(\bar{\cal L}|\bar{\cal N}_{\Sigma})=
\log(W_M(\bar{\cal N}_{\Sigma})), \nonumber
\end{align}
leading to \begin{align} H_{\bar{\cal N}_{\Sigma}|\bar{\cal L}}=
0, \nonumber
\end{align}
\begin{align} H_{\bar{\cal L}|\bar{n}_{\Sigma}}=
\log(W_M(\bar{n}_{\Sigma})), \nonumber
\end{align}
\begin{align} H_{\bar{\cal L}|\bar{\cal N}_{\Sigma}}&=
\sum_{\bar{n}_{\Sigma}\in \{\bar{\cal N}_{\Sigma}\}}p_{\bar{\cal
N}_{\Sigma}}(\bar{\cal N}_{\Sigma})\log(W_M(\bar{n}_{\Sigma}))
\nonumber \\ &= \langle \log(W_M(\bar{\cal N}))\rangle.
\label{condentr}
\end{align}
By (\ref{mutual}) we have
\begin{align} I_{\bar{\cal L};\bar{\cal N}_{\Sigma}}=H_{\bar{\cal L}}
-H_{\bar{\cal L}| \bar{\cal N}_{\Sigma}} = H_{\bar{\cal
N}_{\Sigma}} \geq 0, \label{zeroentropy57}
\end{align}
where the inequality becomes equality only when $\bar{\cal
N}_{\Sigma}$ is non-random. The above inequality says that, while
$H(\bar{\cal L})$ is an unbiased estimator of the Gibbs entropy,
$H(\bar{\cal L}|\bar{\cal N}_{\Sigma})$ is not. Substituting
(\ref{condentr}) in  (\ref{zeroentropy57}) we have
\begin{align}  H_{\bar{\cal N}_{\Sigma}} &
=H_{\bar{\cal L}^N}- \log(N!) +\sum_{l
 \in \{{\cal L}\}}\braket{
\log({\cal N}_{\Sigma}(l)!)} \label{ei1} \\ & =NH_{{\cal L}}-
\log(N!) +\sum_{l
 \in \{{\cal L}\}}\braket{
\log({\cal N}_{\Sigma}(l)!)} \label{ei} \\ &=NH_{\cal L}- \log(N!)
+\sum_{l
 \in \{{\cal L}\}}\braket{
\log({\cal B}(l)!)}\nonumber,\end{align} where ${\cal B}(l)$ is a
binomial random variable with parameter $p=p_{\cal L}(l)$ and $N$
trials, see \cite{me},
\begin{align}\hspace{-0.3cm} \braket{
\log({\cal B}(l))} =\sum_{n=0}^{N}\left(
\begin{array}{c} N \\
n\\
\end{array} \right)
p_{\cal L}^n(l)(1-p_{\cal L}(l))^{N-n} \log(n!),
\label{binomil}\end{align} which  has been recently calculated in
integral form in \cite{mahdi}. Substituting (\ref{shmulti}) in
(\ref{ei1}) we get
\begin{align}
 H_{\bar{\cal N}_{\Sigma}}& \stackrel{\text{Stirling}}{\approx} \log(W_M(\langle
\bar{\cal N})\rangle)- \langle \log(W_M(\bar{\cal N}))\rangle
 \nonumber \\ &  = \sum_l \langle \log({\cal
N}_l)\rangle-\log(\langle{\cal N}_l\rangle) \geq 0,\label{eimult}
\end{align}
where the inequality\footnote{Although it is guaranteed that
$H_{\bar{\cal N}} \geq 0$, in principle the approximation could
become negative.} is obtained by applying to all the terms of the
sum the Jensen inequality for the convex (upward) function
$f(x)=\log(x!)$:
\[\langle f({\cal X}) \rangle \geq f(\langle {\cal X} \rangle).\]
Note that if we put
\[ \langle \log(W(\bar{\cal N}))\rangle \approx \log(W(\langle \bar{\cal
N})\rangle) \]  we get zero entropy of the random occupancy
macrostate! Instead, $H_{\bar{\cal N}_{\Sigma}}$ is approximated
with the quality of the Stirling approximation by the sum of the
Jensen gaps  in the sum (\ref{eimult}).

By (\ref{zeroentropy57}), $H_{\bar{\cal N}_{\Sigma}}$ is equal to
the classical mutual information
 $I_{\bar{\cal L}^N|\bar{\cal N}_{\Sigma}}$ between $\bar{\cal L}^N$ and
$\bar{\cal N}_{\Sigma}$, which can be related to the quantum
entropy as follows. Writing the density operator
(\ref{mixedsystem2}) as
\begin{equation}\hat{\rho}= \sum_{\bar{n}_{\Sigma} \in \{\bar{\cal N}_{\Sigma}\}}p_{\bar{\cal{N}}_{\Sigma}}
(\bar{n}_{\Sigma}) \hat{\rho}(\bar{n}_{\Sigma}),\label{density}
\end{equation}
with
\begin{align}
\hat{\rho}({\bar{n}_{\Sigma}})& = \sum_{\bar{l}^N \in \{\bar{\cal
L}^N\}}p_{\bar{\cal L}^N|\bar{\cal
N}_{\Sigma}}(\bar{l}^N|\bar{n}_{\Sigma})
\ket{\phi(\bar{l}^N)}\bra{\phi(\bar{l}^N)},\label{prep}
\end{align}
we recognize that $H_{\bar{\cal N}_{\Sigma}}$ is equal to the
Holevo upper bound $\chi$ \cite{ikemike} above the accessible
information:
\begin{align}\chi &\stackrel{\text{def}}{=}
S(\hat{\rho})-\sum_{\bar{n}_{\Sigma} \in \{\bar{\cal
N}_{\Sigma}\}} p_{\bar{\cal
N}_{\Sigma}}(\bar{n}_{\Sigma})S(\hat{\rho}({\bar{n}_{\Sigma}}))
\label{vnineq1}
\\ & =H_{\bar{\cal L}^N}- H_{\bar{\cal L}^N|\bar{\cal N}_{\Sigma}}\stackrel{\text{def}}{=}
I_{\bar{\cal L}^N|\bar{\cal N}_{\Sigma}} \label{ort}
\\ & =H_{\bar{\cal N}_{\Sigma}}, \label{ze}
\end{align}
where, by the diagonal nature of the density operator
$\hat{\rho}({\bar{n}_{\Sigma}})$,
\begin{align}
S(\hat{\rho}({\bar{n}_{\Sigma}}))=H_{\bar{\cal
L}^N|\bar{n}_{\Sigma}},\nonumber
\end{align}
\begin{align}
\sum_{\bar{n}_{\Sigma} \in \{\bar{\cal N}_{\Sigma}\}} p_{\bar{\cal
N}_{\Sigma}}(\bar{n}_{\Sigma})S(\hat{\rho}({\bar{n}_{\Sigma}}))=
H_{\bar{\cal L}^N|\bar{\cal N}_{\Sigma}}. \nonumber
\end{align}
Note that factorization of $\hat{\rho}$ in (\ref{seppart}) does
not imply factorization of $\hat{\rho}(\bar{n}_{\Sigma})$,
basically because factorization of $p_{\bar{\cal L}^N}$ does not
imply factorization of $p_{\bar{\cal L}^N|\bar{\cal N}_{\Sigma}}$.
Therefore, while the particles of the system represented by
$\hat{\rho}$ are not entangled between them, the particles of the
preparation $\hat{\rho}(\bar{n}_{\Sigma})$ may be entangled
between them. The deep connection between information and
thermodynamics has been enlightened by the paper of Landauer
\cite{landauer}, that showed the equivalence between heat and
logical information. Landauer's result is today proved
experimentally \cite{landdemonstration} and it is widely accepted
that (quantum) thermodynamics can be treated by (quantum)
information-theoretic tools, see for instance the tutorial paper
\cite{plenio}.

A comment is in order about the interpretation of the word {\em
accessible} between {\em Holevo} and {\em information} in this
specific context. It is indistinguishability that makes part of
the entropy $H_{\bar{\cal L}^N}$ of distinguishable particles {\em
non accessible}. Actually, from equations (\ref{vnineq1}) and
(\ref{ort}) is clear that lack of indexing (hence
indistinguishability) makes impossible to know which among the
$W(\bar{\cal N}_{\Sigma})$ microstates is visited by the system
whose macrostate is $\bar{\cal N}_{\Sigma}$, hence lack of
indexing prevents the access to the corresponding
$\log(W(\bar{\cal N}_{\Sigma}))$ nats of information, that, in the
average, are just $H_{\bar{\cal L}^N|\bar{\cal N}_{\Sigma}}$ nats.

Finally, note that, using Stirling's formula and
\begin{equation}\lim_{N \rightarrow \infty}\frac{{\cal
N}_{\Sigma}(l)}{N}=p_{\cal L}(l),\nonumber
\end{equation} it is straightforward to manipulate the random
$W_M(\bar{\cal N}_{\Sigma})$ to show that
\begin{align} \lim_{N
\rightarrow \infty} \frac{ \log(W_M(\bar{\cal N}_{\Sigma}))}{N}
 &= H_{{\cal L}},\label{aepcons}
\end{align}
\begin{align} \lim_{N
\rightarrow \infty} \frac{ H_{\bar{\cal L}^N}-H_{\bar{\cal
L}^N|\bar{\cal N}_{\Sigma}}}{N}
 &= \lim_{N
\rightarrow \infty} \frac{ H_{\bar{\cal N}_{\Sigma}}}{N}
 =0.\label{lack}
\end{align}
Hence, for $N \rightarrow \infty$, the randomness of $\bar{\cal
N}_{\Sigma}$ impacts only the conditional entropy, not the
conditional entropy per particle, which is actually equal to the
entropy per particle. Equation (\ref{aepcons}) expresses what in
information theory is called Asymptotic Equipartition Property
(AEP), which stands at the basis of the information-theoretic
typicality. Appendix C reports a discussion of the celebrated
$\log(W)$ formula at the light of the AEP and of the
information-theoretic typicality.

\subsection*{Other macrostates}

Other random macrostates than the occupancy macrostate can be
obtained by the union of all the microstates associated to the
same value of the random sum
\[{\cal M}_{\Sigma}=\sum_{i=1}^{N}m({\cal L}_i),\]
where $m(\cdot)$ is a deterministic function of the quantum
number. For instance, the random energy macrostate, which is
characterized by the total energy ${\cal E}_{\Sigma}$, is the
union of all the microstates whose total energy is
\[{\cal E}_{\Sigma}=\sum_{i=1}^{N}\epsilon({\cal L}_i).\]

The random occupancy macrostate is the most detailed random
macrostate because from it any other random macrostate can be
derived by writing
\begin{equation}{\cal M}_{\Sigma}=\sum_{i=1}^{N}m({\cal L}_{i})=
\sum_{l \in \{{\cal L}\}}{\cal N}_{\Sigma}(l)
m(l),\label{randomex}
\end{equation} where the second equality can be seen
as the random version of the expectation of $m({\cal L})$, with
$\bar{\cal N}_{\Sigma}$ in the role of ''random'' probability
distribution multiplied by the number of particles. The
probability of a macrostate is the probability that comes out a
microstate that belongs to the macrostate, hence the probability
of the macrostate can be identified with the probability of the
sum:
\begin{align}
Pr(\bar{\cal L} \in \{\bar{l}({m}_{\Sigma})\})=p_{{\cal
M}_{\Sigma}}({m_{\Sigma}}).
 \nonumber
\end{align}
The set of vectors belonging to $\{\bar{l}({m}_{\Sigma})\}$ can be
partitioned into non-overlapping subsets of the type
(\ref{permset}), hence the probability distribution of the
macrostate is
\begin{equation}p_{{\cal
M}_{\Sigma}}(m_{\Sigma})=\sum_{\bar{n}_{\Sigma}: \sum_{l \in
\{{\cal L}\}}{n}_{\Sigma}(l)m(l)=m_{\Sigma}}p_{\bar{\cal
N}_{\Sigma}}(\bar{n}_{\Sigma}),\label{pmacro}
\end{equation}
and the number of microstates belonging to the macrostate is
\begin{equation}W(m_{\Sigma})=\sum_{\bar{n}_{\Sigma}: \sum_{l
\in \{{\cal
L}\}}{n}_{\Sigma}(l)m(l)=m_{\Sigma}}W_M(\bar{n}_{\Sigma}).\label{wsumenergy}
\end{equation}

Any generic random macrostate $\bar{\cal M}_{\Sigma}$ is known
given $\bar{\cal N}_{\Sigma}$, therefore $H_{\bar{\cal
M}_{\Sigma}|\bar{\cal N}_{\Sigma}}=0$, leading to
\[I_{\bar{\cal N}_{\Sigma};\bar{\cal M}_{\Sigma}}=H_{\bar{\cal N}_{\Sigma}}-
H_{\bar{\cal N}_{\Sigma}|\bar{\cal M}_{\Sigma}}= H_{\bar{\cal
M}_{\Sigma}}-H_{\bar{\cal M}_{\Sigma}|\bar{\cal
N}_{\Sigma}}=H_{\bar{\cal M}_{\Sigma}},\]
\begin{align} I_{\bar{\cal L};\bar{\cal M}_{\Sigma}}= H_{\bar{\cal M}_{\Sigma}}
=I_{\bar{\cal N}_{\Sigma};\bar{\cal M}_{\Sigma}}. \nonumber
\end{align}Since
\[H_{\bar{\cal N}_{\Sigma}} \geq
H_{\bar{\cal M}_{\Sigma}},\] also the entropy per particle of any
macrostate becomes vanishingly small as $N \rightarrow \infty$.

\section{Thermal equilibrium}

Suppose that the system is at the thermal equilibrium with a heat
bath at known temperature $T$ Kelvin degrees. Even if, besides the
temperature and the number of particles, other constraints are
imposed, e.g., in the case of a gas, the volume occupied by the
particles, the constraints can be not enough to uniquely define
the probability distribution $p_{\cal L}$. To find this
distribution, Jaynes proposed in \cite{maxent} to maximize the
entropy of the microstate $H_{\bar{\cal L}}$ under the temperature
constraint. This is the famous {\em maxent} principle, which
become an universally accepted standard after Jaynes.  In our
approach, the entropy to be maximized should be the entropy of the
occupancy macrostate $H_{\bar{\cal N}_{\Sigma}}$. However, the sum
appearing in (\ref{ei1}) makes maximization difficult. If the sum
in (\ref{ei1}) is neglected, the result of the maximization of
$H_{\bar{\cal N}_{\Sigma}}$ is the same as that of the
maximization of $H_{\bar{\cal L}}$, because the term $\log(N!)$ of
(\ref{ei1}) is independent of the probability distribution. Since
the $N$ entries of the random vector $\bar{\cal L}^N$ are assumed
to be i.i.d., the entropy $H_{\bar{\cal L}^N}$ is $N$ times the
entropy of one particle, therefore in the maximization we can
consider the one-particle entropy. Maximization of $H_{{\cal L}}$
can be worked out by the Lagrange multipliers method, the
Lagrangian being
\begin{align}
\Lambda &=-\sum_{l \in \{{\cal L}\}}p_{\cal L}(l)
\left(\log(p_{\cal L}(l)) +\beta \epsilon(l) +
\alpha \right), \nonumber
\end{align}
where $\beta$ and $\alpha$ are the Lagrange multipliers. The
Lagrange multiplier $\alpha$ is found by imposing the constraint
\[\sum_lp_{\cal L}(l)=1.\]
The Lagrange multiplier $\beta$ is found by imposing that
$k_BH_{\cal L}$ is the thermodynamical (Clausius) entropy per
particle, the quantity usually denoted $S/N$, that is by imposing
\begin{equation}\frac{\partial H_{\cal L}}{\partial \mu_{\cal
E}}=\frac{1}{k_BT}.\label{clausius}
\end{equation}

The partial derivatives of the Lagrangian are
\begin{align}
\frac{\partial \Lambda}{\partial p_{\cal L}(l)}&=- \log(p_{\cal
L}(l)) -1-\beta \epsilon(l) -\alpha,  \ \ \forall \ l \in \{{\cal
L}\},\nonumber
\end{align}
the solution is therefore
\begin{equation}
p_{\cal L}(l)=e^{-\beta \epsilon(l)-\alpha-1}, \label{boltz5}
\end{equation}
where $\alpha$ is such that
\begin{align}
\sum_{l \in \{{\cal L}\}}p_{\cal L}(l)=\sum_{l \in \{{\cal
L}\}}e^{-\beta \epsilon(l)-\alpha-1}=1, \nonumber
\end{align}
hence
\begin{align}
e^{\alpha+1}=\sum_{l \in \{{\cal L}\}}e^{-\beta \epsilon(l)}.
\nonumber
\end{align}
The above sum takes the name of (one-particle) {\em canonical
partition function} and is usually indicated with the symbol $Z$:
\begin{align}
Z=e^{\alpha+1}=\sum_{l \in \{{\cal L}\}}e^{-\beta \epsilon(l)}.
\label{partition}
\end{align}
Substituting (\ref{partition}) in (\ref{boltz5}) one gets the
Boltzmann distribution in the familiar form
\begin{equation}
p_{\cal L}(l)=\frac{1}{Z}e^{-\beta \epsilon(l)}. \label{boltz}
\end{equation}
The probability distribution of the discrete energy is
\[p_{\cal E}(\epsilon)=\frac{g_{\epsilon}}{Z}e^{-\beta
\epsilon},\] where $g_{\epsilon}$ is the degeneracy of the energy
eigenvalue $\epsilon$, that is the number of distinct eigenstates
with the same energy eigenvalue $\epsilon$. The Boltzmann
distribution (\ref{boltz}) can now be used to find the
one-particle random entropy and its expectation:
\begin{align}
H({\cal L})&=\log(Ze^{\beta \epsilon({\cal L})}), \nonumber
\end{align}
\begin{align}
H_{{\cal L}}= \sum_{l \in \{{\cal L}\}}(Ze^{\beta
\epsilon(l)})^{-1} \log(Ze^{\beta \epsilon(l)}).
\label{boltzentropy}
\end{align}
We can manipulate the above two formulas as follows:
\begin{align}
H({\cal L})&=\log(Z) + \beta \epsilon({\cal L}), \nonumber
\end{align}
\begin{align}
H_{{\cal L}}= \log(Z)+\beta \mu_{\cal E}, \label{boltzentropy5}
\end{align}
where
\begin{align}
\mu_{\cal E}=\frac{1}{Z}\sum_{l \in \{{\cal L}\}}\epsilon(l)
e^{-\beta \epsilon(l)} \label{beta}
\end{align}
is the expect energy per particle, a quantity often called $U/N$
in standard textbooks, where $U$ is system's internal energy. If
desired, $\beta$ can be found as a function of $\mu_{\cal E}$ by
solving for $\beta$ the transcendental equation (\ref{beta}).
However, when the entropy-$\beta$ relation is the concern, there
is no need of expressing $\beta$ as a function of the expected
energy per particle, it is enough to use directly
(\ref{boltzentropy}), or, equivalently, to substitute (\ref{beta})
in (\ref{boltzentropy5}).

The temperature constraint can be imposed by observing that
\begin{align}\frac{\partial \log(Z)}{\partial
\beta}&=\frac{\partial \log(Z)}{\partial Z} \frac{\partial
Z}{\partial \beta}\nonumber \\ &= \frac{1}{Z} \left(-\sum_{l \in
\{{\cal L}\}} \epsilon e^{-\beta \epsilon(l)} \right)\nonumber
\\ &=-\mu_{\cal E},\nonumber
\end{align}
henceforth
\begin{align}\frac{\partial H_{{\cal L}}}{\partial
\mu_{\cal E}}&=\frac{\partial \log(Z)}{\partial \mu_{\cal E}}+
\beta +\mu_{\cal E} \frac{\partial \beta}{\partial \mu_{\cal E}}
\nonumber
\\ &=\frac{\partial \log(Z)}{\partial \beta}
\frac{\partial \beta}{\partial \mu_{\cal E}} + \beta + \mu_{\cal
E} \frac{\partial \beta}{\partial \mu_{\cal E}} \nonumber
\\ &= \beta=\frac{1}{k_BT}.\nonumber
\end{align}
where the last equality is the constraint (\ref{clausius}). All
the classical thermodynamics of systems at the equilibrium can be
obtained from the identification of $\beta$ with the inverse
temperature and of $k_B H_{\cal L}$ with the Clausius entropy per
particle. For instance, since
\begin{align}\frac{\partial T \log(Z)}{\partial
T}&=H_{\cal L}, \nonumber
\end{align}
from (\ref{boltzentropy5}) we have that the Helmotz free energy
per particle $F/N$ is
\[\frac{F}{N}\stackrel{\text{def}}{=}-k_BT \log(Z)=\mu_{\cal E}-k_BTH_{\cal L},\]
an equation that in the standard textbook notation reads
\[F=U-TS.\]

The probability distribution of microstates is
\begin{align}
p_{\bar{\cal L}}(\bar{l})&=\prod_{i=1}^{N}p_{{\cal
L}}(l_i)\nonumber
\\ & =\prod_{i=1}^{N}\frac{e^{-\beta \epsilon(l_i)}}{Z}
\nonumber \\ & =\frac{e^{-\beta
\sum_{i=1}^{N}\epsilon(l_i)}}{Z^N}. \label{boltz2} \end{align} For
the random system entropy, from (\ref{boltz2}) we have
\begin{equation}
H(\bar{\cal L})=N\log(Z)+\beta\sum_{i=1}^{N}\epsilon({\cal
L}_i),\nonumber
\end{equation}
whose expectation is
\begin{equation}
H_{\bar{\cal L}}= N (\log(Z)+\beta \mu_{\cal E}). \nonumber
\end{equation}
The Boltzmann distribution (\ref{boltz2}) for systems at the
thermal equilibrium is one of the central results of thermal
physics and statistical mechanics. A derivation very similar to
ours but based on the constrained maximization of the multinomial
coefficient can be found, for instance, in 18.1 of \cite{sek} and
in 3.2 of \cite{pathria}.  Another derivation of (\ref{boltz2})
can be obtained by separating the system from the heath bath and
by assuming equiprobability of microstates of the heath bath after
separation, see for instance 19.1 of \cite{sek}, 3.1 of
\cite{pathria}.

\subsection*{Quantum harmonic oscillator at the thermal equilibrium}

For the quantum harmonic oscillator (\ref{pib0}) at the thermal
equilibrium, the probability distribution of the energy levels,
the entropy, the partition function and the energy-temperature
ratio can be found in closed form by calculating the two sums in
(\ref{partition}) and (\ref{beta}). To our knowledge, calculus of
(\ref{partition}) and (\ref{beta}) dates back to Einstein
\cite{einstein1907,einstein1909} and was applied to the entropy in
Section II of \cite{gordon}. The partition function
(\ref{partition}) turns out to be a geometric series, whose
calculation gives\footnote{As in 3.8 of \cite{pathria}, for
completeness we consider the energy of the ground state in the
calculations. However, as observed in 16.3 of \cite{sek} and in
\cite{gordon}, adding the energy of the ground state produces only
a shift that does not impact neither the probability distribution
nor the entropy.}
\begin{equation}
Z=e^{-\frac{\beta\omega \hbar}{2}}\sum_{l=0}^{\infty}e^{-\beta l
\omega \hbar}= \frac{e^{\frac{\beta\omega \hbar}{2}}}{e^{\beta
\omega \hbar}-1}. \label{sumz}
\end{equation}
Using (\ref{sumz}) in (\ref{boltz}), we see that the probability
distribution of the random quantum number ${\cal L}$ of
(\ref{pib0}) is the geometric distribution:
\begin{align}
p_{{\cal L}}(l)&=(1-p)p^{l}, \ \ \ l=0,1, \cdots, \label{gordon}
\end{align}
with
\begin{align}
p=e^{-\beta \omega \hbar}=1-\frac{e^{-\frac{\beta\omega
\hbar}{2}}}{Z}, \label{geobasis}
\end{align}
see 7.5.6 of \cite{kly} for the derivation of (\ref{gordon}) in a
full quantum-mechanical approach. The mean value and the variance
of the geometric distribution are known to be\footnote{Einstein
calculates the expected energy (hence the sum (\ref{beta})) as
\begin{align}
\mu_{\cal E}&=\frac{\omega \hbar}{2}+\frac{\omega \hbar
e^{-\frac{\beta l \omega \hbar}{2}}}{Z} \sum_{l=0}^{\infty}l
e^{-\beta l \omega \hbar} \nonumber
\\ &=\frac{\omega \hbar}{2}+\frac{\omega \hbar e^{-\frac{\beta l \omega \hbar}{2}}}{Z}\frac{ e^{\beta
\omega \hbar}}{(e^{\beta \omega \hbar}-1)^2} \nonumber
\\ &=\frac{\omega \hbar}{2}+\frac{
\omega \hbar}{e^{\beta \omega \hbar}-1}, \label{expe}
\end{align}
which, together with
\begin{equation}
\mu_{\cal L}=\frac{\mu_{\cal E}}{\omega
\hbar}-\frac{1}{2},\nonumber
\end{equation}
gives (\ref{mul}).}
\begin{align}
\mu_{\cal L}&= \frac{p}{1-p}= \frac{e^{-\beta \omega
\hbar}}{1-e^{-\beta \omega \hbar}}= \frac{1}{e^{\beta \omega
\hbar}-1}, \label{mul}
\end{align}
\begin{align}
\sigma^2_{\cal L}&=\frac{p}{(1-p)^2}= \mu_{\cal L}(1+\mu_{\cal
L})=\frac{e^{-\beta \omega \hbar}}{(1-e^{-\beta \omega \hbar})^2}.
\nonumber
\end{align}
At high temperature, where
\begin{align}e^{\pm \beta \omega \hbar} \approx 1 \pm \beta \omega \hbar,
\label{ht}\end{align} we have
\[\sigma_{\cal L} \approx \mu_{\cal L}
\approx \frac{1}{\beta \omega \hbar}. \]

Taking the mean value as a parameter, the geometric distribution
can be expressed in the form
\begin{align} p_{\cal L}(l)&=\frac{1}{1+\mu_{\cal L}}\left(
\frac{\mu_{\cal L}}{1+\mu_{\cal L}}\right)^{l}, \ \ \ l = 0,1,
\cdots, \nonumber
\end{align}
leading, for the random entropy and for the entropy, to
\begin{align}
H({\cal L})&=\log(1+\mu_{\cal L})+{\cal L}\log(1+\mu_{\cal
L}^{-1}) ,\nonumber
\end{align}
\begin{align}
H_{{\cal L}}&=\log(1+\mu_{\cal L})+\mu_{\cal L}\log(1+\mu_{\cal
L}^{-1}) \label{eqho} \\ &= (1+\mu_{\cal L})\log(1+\mu_{\cal L})-
\mu_{\cal L}\log(\mu_{\cal L}),\nonumber
\end{align}
where the last line is the form used by Planck in the analysis of
the blackbody radiation spectrum, but for Planck the energy of the
ground state is zero, so after formula (10) of his 1901 paper
\cite{planck} he puts $\mu_{\cal E}/\omega \hbar$ in place of
$\mu_{\cal L}$. Substituting the last term of (\ref{mul}) in place
of $\mu_{\cal L}$ in the above equation, the entropy-temperature
relation results
\begin{align}
H_{{\cal L}}&=\log\left(\frac{1}{1-e^{-\beta \omega
\hbar}}\right)+\frac{\beta \omega \hbar}{e^{\beta \omega
\hbar}-1}.\label{gordone}
\end{align}

Substituting (\ref{ht}) in the above equation, at high temperature
we have
\begin{align}
H_{{\cal L}}&\approx \log\left(\frac{1}{1-e^{-\beta \omega
\hbar}}\right)+1,\label{gordoneht}
\end{align}
which, compared to (\ref{gordone}), shows that
\[\beta \mu_{\cal E}=\frac{\omega \hbar}{2}+ \frac{\beta \omega \hbar}{e^{\beta \omega
\hbar}-1}\approx \frac{\beta \omega \hbar}{e^{\beta \omega
\hbar}-1} \approx 1.\] Equivalently, since at high temperature
$\mu_{\cal L}$ is large, we can put
\[\log(1+\mu_{\cal L}^{-1})\approx \mu_{\cal L}^{-1}\]
in (\ref{eqho}), leading to
\begin{align}
H_{{\cal L}}&\approx \log(1+\mu_{\cal L})+1,\nonumber
\end{align}
which, on substitution of (\ref{mul}), is (\ref{gordoneht}). See
appendix D for a system of $N$ monochromatic quantum harmonic
oscillators.

\subsection*{Particle in a box at the thermal equilibrium}

 The Boltzmann distribution for the particle in a box with aperiodic boundary conditions (\ref{pib})
 is
\begin{align}p_{\cal L}(l)&=Z^{-1}e^{-\beta l^2\frac{\pi^2
\hbar^2}{2mL^2}}, \ \ l= 1,2, \cdots.\end{align} Here the
partition function cannot be calculated in closed form, even if
numerical evaluation is possible.

At high temperature, where quantization effects become negligible,
sums are approximated to integrals, discrete random variables are
approximated to dense random variables, and the gas behaves
approximately as a classical gas made by hard balls. Specifically,
the sums (\ref{partition}) and (\ref{beta}) can be approximated to
the discretization with unit step of the following two integrals
\begin{equation}
Z \approx \int_{l=0}^{\infty}e^{- \beta l^2 \frac{ \pi^2
\hbar^2}{2 m  L^2}} dl= \sqrt{\frac{ m  L^2}{2 \beta \pi
\hbar^2}}, \label{betaboxint1}
\end{equation}
and
\begin{equation}
A=\frac{\pi^2 \hbar^2}{2 m  L^2}\int_{l=0}^{\infty}l^2e^{-\beta
l^2 \frac{ \pi^2 \hbar^2}{2 m  L^2}} dl= \sqrt{\frac{  m  L^2}
{8\beta^3 \pi \hbar^2}}, \nonumber
\end{equation}
respectively, leading to
\[\mu_{{\cal E}} \approx \frac{A}{Z} \approx \frac{1}{2 \beta},\]
\begin{align}
H_{{\cal L}}&=\log(Z)+ \beta \mu_{{\cal E}} \nonumber
\\ & \approx \frac{1}{2}\left(\log \left(\frac{ m L^2 }{2  \beta
\pi \hbar^2}\right)+1 \right) \label{secondappr}.
\end{align}

The probability density function of dense (continuous) random
${\cal L}$ is obtained by looking at the quantization rule
appearing in the exponent of the Boltzmann distribution as at a
continuous function of continuous variable. The normalization of
the dense distribution is just the division by the integral of
equation (\ref{betaboxint1}), leading to
\[f_{\cal L}(l)=\sqrt{\frac{2 \beta \pi \hbar^2}{ m  L^2}}
e^{- \frac{\beta \pi^2 \hbar^2}{2 m  L^2 } l^2}, \ \ l>0,\] where
$f_{\cal X}(\cdot)$ indicates the probability density function of
the continuous random variable ${\cal X}$. The probability density
function of the continuous random energy, that again we call
${\cal E}$, is found by making the following change of random
variable from the continuous energy level ${\cal L}$ to the
continuous energy ${\cal E}$:
\[{\cal E}=\frac{ \pi^2 \hbar^2}{2 m L^2}{\cal L}^2,\]
hence
\[{\cal L}=\sqrt{\frac{2 m  L^2 {\cal E}}{ \pi^2 \hbar^2}}.\]
Calling $\epsilon$ the continuous variable that spans the support
of ${\cal E}$, the change of random variable provides us with
\begin{align}f_{\cal E}(\epsilon)&=f_{\cal L}\left(\sqrt{\frac{2 m  L^2 \epsilon}
{ \pi^2 \hbar^2}}\right) \cdot \frac{d}{d \epsilon}\sqrt{\frac{2 m
L^2 \epsilon}{ \pi^2 \hbar^2}}\nonumber \\ &= \sqrt{\frac{2\beta
\pi \hbar^2}{m  L^2}} e^{-\beta \epsilon} \cdot \sqrt{\frac{ m
L^2}{2 \pi^2 \hbar^2 \epsilon}}\nonumber
\\ &= \sqrt{\frac{\beta }{\pi \epsilon}} e^{-\beta \epsilon}\nonumber \\
&=\frac{1 }{\Gamma(0.5)}\sqrt{\frac{\beta }{\epsilon}} e^{-\beta
\epsilon}, \ \ \epsilon >0, \nonumber
\end{align}
that is a Gamma distribution $\Gamma(k, \theta)$ with shape
parameter $k=0.5$ and scale parameter $\theta=\beta^{-1}$, or,
equivalently, a $\chi^2$ distribution with one degree of freedom
and mean value $(2 \beta)^{-1}$.  The probability density function
of momentum of the classical one-dimensional gas particle at the
thermal equilibrium is in fact known to be normal with zero mean
value and variance $m \beta^{-1}$, therefore, since the energy is
wholly kinetic, the probability density function of the continuous
energy must be $\chi^2$ with one degree of freedom and mean value
$(2 \beta)^{-1}$, as it actually is.

\section{Ideal gas in a container}

The pseudo-quantum model for a particle of an ideal gas in a
container is that of the particle in a box, whose energy levels
are given in (\ref{pib}). As we will see, through the quantization
rule (\ref{pib}) the volume of the container will be imported in
the probability distribution of energy levels and, as a
consequence, in the entropy formula. In the following, we take as
a deterministic constraint the Boltzmann distribution for the
energy of gas' particles, even if this could be not the
entropy-maximizing distribution, because entropy maximization
should take into account also the third term of (\ref{ei}), which
is important at low temperature. However, this maximization is out
of the scope of the present paper.

When the $N$ particles have $D$ degrees of freedom, the energy
level becomes a $D$-dimensional vector $\bar{l}^D=(l_{1},l_{2},
\cdots, l_{D})$, the energy level of the $i$-th particle is the
$D$-dimensional vector $\bar{l}_i^D=(l_{i,1},l_{i,2}, \cdots,
l_{i,D})$ and the occupancy number is the number of particles that
are found in a vector of energy levels, hence the occupancy number
is
\[{\cal N}_{\Sigma}(\bar{l}^D)=
\sum_{i=1}^N\prod_{d=1}^D\delta({\cal L}_{i,d}-l_d),
\]
where ${\cal L}_{i,d}$ is the random energy level of the $i$-th
particle in the $d$-th dimension and
\[\sum_{i=1}^N{\cal N}_{\Sigma}(\bar{l}^D)=N.\]
With a $D$-dimensional box with sides all of length $L$ and
independency between the energy levels of the individual
dimensions, that is
\[p_{{\bar{\cal
L}^D}}(\bar{l}^D)=\prod_{d=1}^D p_{{\cal L}}(l_d),\] the entropy
of the system of $N$ non-indexed particles is
\begin{align}
H_{\bar{\cal N}_{\Sigma}}&=NDH_{\cal L}-\log(N!)+\sum_{\bar{l}^D
\in \{\bar{\cal
L}^D\}}\braket{\log({\cal N}_{\Sigma}(\bar{l}^D)!)} \label{3dexact} \\
& \approx NDH_{\cal L}-\log(N!) \label{twoterms} \\
& \approx \frac{ND}{2}\left(\log \left(\frac{ m
 L^2 }{2 \beta \pi \hbar^2}\right)+1 \right)
 -\log(N!)\label{3dapprox1} \\
& = N\left(\log \left( L^D\left(\frac{m }{2 \beta \pi
\hbar^2}\right)^{\frac{D}{2}}\right)+\frac{D}{2} \right)
-\log(N!)\nonumber
\\
& \stackrel{\text{Stirling}}{\approx} N\left(\log \left(
\frac{L^D}{N}\left(\frac{m }{2 \beta \pi \hbar^2
}\right)^{\frac{D}{2}}\right)+\frac{D+2}{2} \right),
\label{3dapprox3}
\end{align}
The two approximations (\ref{twoterms}) and (\ref{3dapprox1}) rule
out two different quantum effects that become negligible at high
temperature. The quantum effect neglected in (\ref{twoterms}),
which impacts the conditional entropy $H_{\bar{\cal L}|\bar{\cal
N}_{\Sigma}}$, is represented by the multiple sum appearing in
(\ref{3dexact}). This multiple sum accounts for the probability
that two or more particles are found in the same vector of energy
levels, hence its nature is clearly quantistic. The approximation
(\ref{3dapprox1}), which impacts the entropy $H_{{\cal L}}$, is
the approximation of the discrete energy to a dense variable that
has been discussed before, see (\ref{secondappr}). Stirling's
approximation finally leads to the $D$-dimensional version of the
Sackur-Tetrode entropy formula (\ref{3dapprox3}). At low
temperature, the entropy $H_{\cal L}$ cannot be calculated in
closed form, but numerical evaluation is possible, as well as it
is possible numerical evaluation of the sum in the first line of
(\ref{3dexact}).\footnote{Note that the sum in the first line of
(\ref{3dexact}), whose generic term is
\begin{align}& \langle\log({\cal N}_{\Sigma}(\bar{l}^D)!)\rangle=
\sum_{n=0}^{N}\left(
\begin{array}{c} N \\
n\\
\end{array} \right)
(p_{\bar{\cal L}^D}(\bar{l}^D))^n(1-p_{\bar{\cal
L}^D}(\bar{l}^D))^{N-n} \log(n!), \nonumber\end{align} is a
multiple sum over $D$ indexes, hence its numerical evaluation can
become demanding for large $D$. The complexity of the evaluation
can be mitigated by suitably truncating the set $\{\bar{{\cal
L}}^D\}$ to those vectors whose probability is non-negligible. At
low temperature, vectors with non-negligible probability have
entries that take their values with high probability in the first
few one-dimensional energy levels, while at high temperature this
quantum effect can be neglected. In practice, this means that we
can take $|\bar{\cal L}^D|=|{\cal L}|^D$ with $|{\cal L}|$ that
becomes smaller and smaller as $D$ increase.}

\begin{figure}[!h]
\vspace*{.2cm}
    \centering
    \includegraphics[width=.52\textwidth]{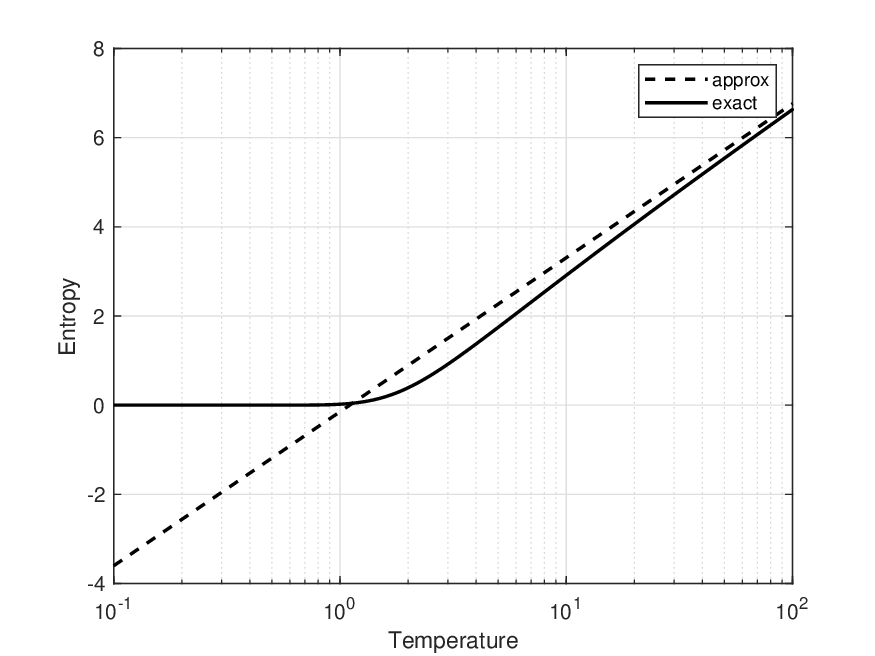}
    \caption{
    Solid line: numerical evaluation of $3H_{{\cal L}}$.
    Dashed line: approximation to $3H_{\cal L}$ obtained by
    (\ref{secondappr}).
    }
    \label{fig:approx}
    \vspace*{.3cm}
\end{figure}

\begin{figure}[!h]
\vspace*{.2cm}
    \centering
    \includegraphics[width=.52\textwidth]{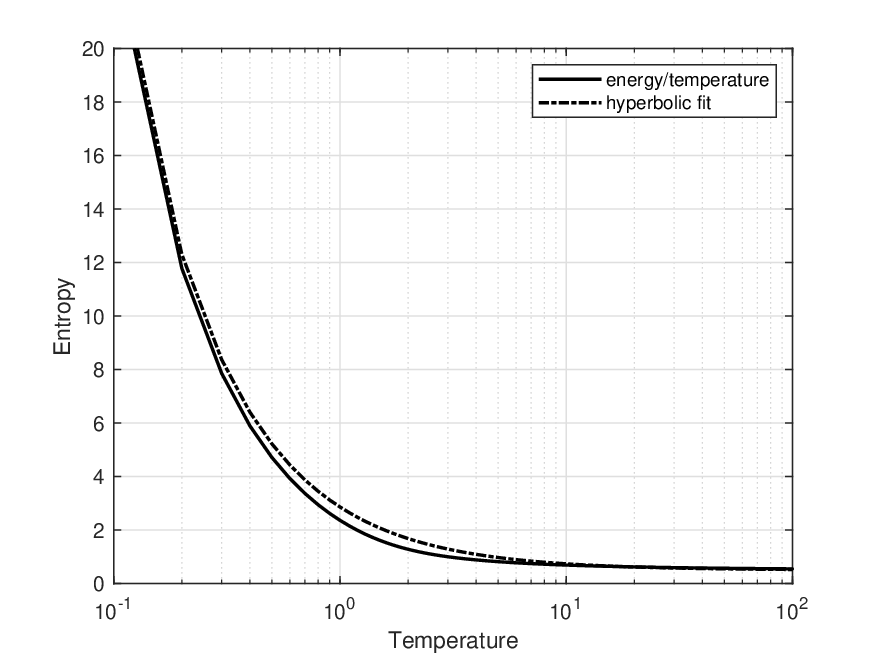}
    \caption{
    Solid line: energy/temperature contribution $\beta \mu_{{\cal E}}$
    to the entropy per particle per dimension.
    Dash-dotted line: hyperbolic fit (\ref{hyfit}).
    }
    \label{fig:hyp}
    \vspace*{.3cm}
\end{figure}

\begin{figure}[!h]
\vspace*{.2cm}
    \centering
    \includegraphics[width=.52\textwidth]{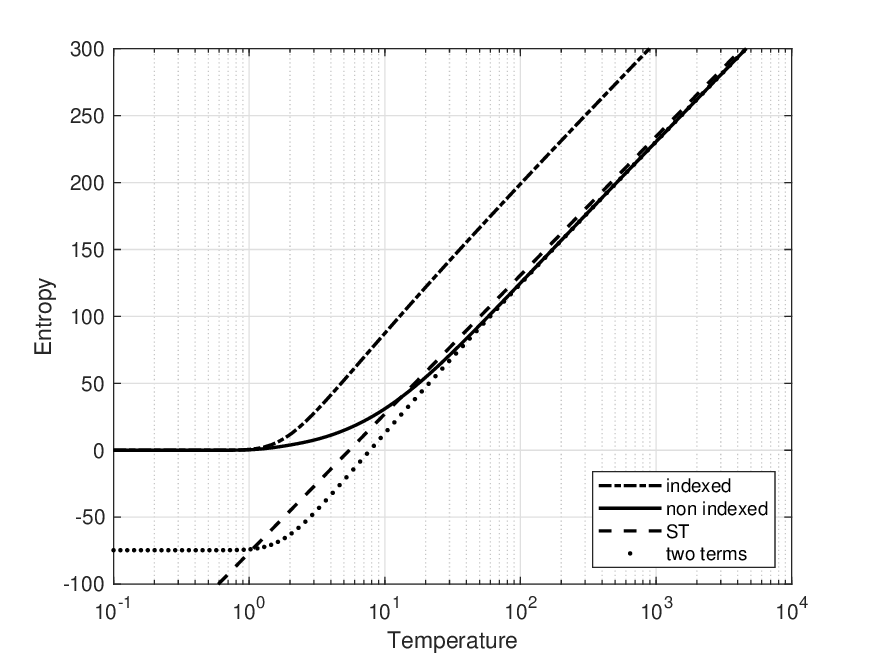}
    \caption{Results for various entropies for particles with three degrees of freedom ($D=3$).
    Dash-dotted line: entropy of indexed particles, $3NH_{\cal L}$ in (\ref{3dexact}).
    Solid line: entropy of non-indexed particles, $H_{\bar{\cal N}_{\Sigma}}$ in (\ref{3dexact}).
    Dotted line: equation (\ref{twoterms}). Dashed line: Sackur-Tetrode entropy (\ref{3dapprox3}).}
    \label{fig:piab30}
    \vspace*{.3cm}
\end{figure}

\begin{figure}[!h]
\vspace*{.2cm}
    \centering
    \includegraphics[width=.52\textwidth]{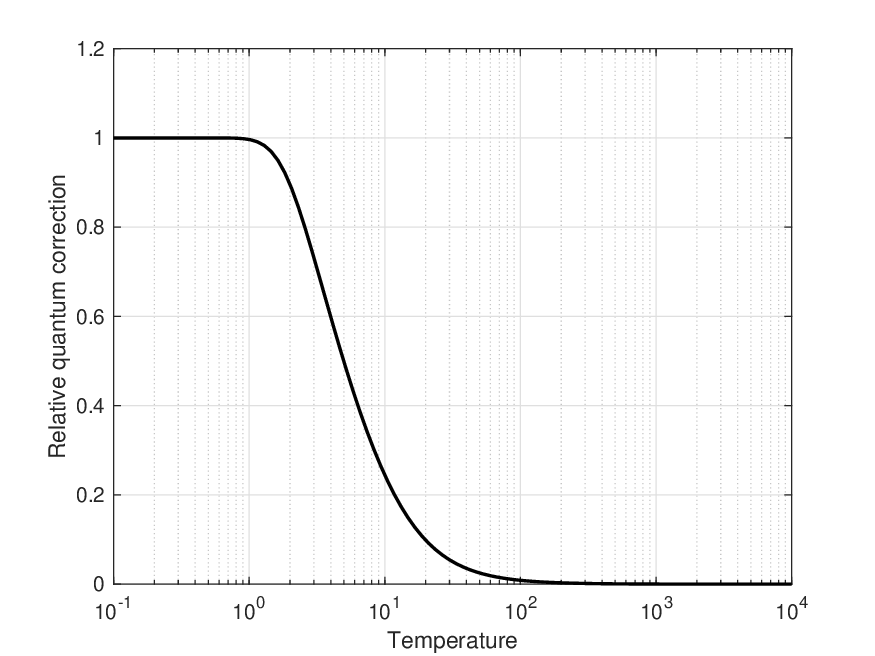}
    \caption{
    Relative quantum correction, third term of (\ref{3dexact}) divided by $\log(N!)$.}
    \label{fig:corr30}
    \vspace*{.3cm}
\end{figure}

The results of Figs. \ref{fig:approx}, \ref{fig:hyp},
\ref{fig:piab30},
 \ref{fig:corr30}, report the entropy (or terms that
contribute to the entropy) in $k_B$ units versus temperature in
Kelvin degrees. All the results are obtained for volume of the
cubic box $V=10^{-27}$ cubic meters and mass of one particle
$m=1.67 \cdot 10^{-27}$ kilograms. With these parameters, the
energy of the one-dimensional ground state is
\[\epsilon(1)=\frac{\pi^2 \hbar^2}{2 m L^2}=3.25 \cdot 10^{-23}.\]
In Figs. \ref{fig:approx}, \ref{fig:hyp}, \ref{fig:piab30}, the
size of the set of energy level has been truncated to $|{\cal
L}|=1000$, having verified that the probability of ${\cal L}>1000$
is so small that it does not impact the graphs in the entire range
of temperature. We hasten to point out that the graphs reported in
the following are intended to be only a numerical exercise based
on the formulas presented in this paper, therefore they will
describe the behavior of a real physical system only inside the
range of validity of the assumptions that we made.

 Fig. \ref{fig:approx} shows the
approximation (\ref{secondappr}) and the numerical evaluation of
the entropy $3H_{\cal L}$ of one particle in three dimensions. The
energy/temperature term $ \beta \mu_{\cal E}$ is reported in Fig.
\ref{fig:hyp}. At $T=1$ Kelvin this term is approximately equal to
2.37, which corresponds to the following expected energy per
dimension
\[\mu_{\cal E}|_{T=1} =
1.005 \cdot \epsilon(1).\] Since the mean value of the energy per
degree of freedom is lower bounded by $\epsilon(1)$ and since the
lower bound is virtually reached at 1 Kelvin, at temperature below
1 Kelvin the energy/temperature term $\beta \mu_{\cal E}$ is
forced to grow hyperbolically. In passing, we observe that the
energy/temperature term $\beta \mu_{\cal E}$ strongly resembles a
hyperbole in the entire range of temperature. The heuristic
hyperbolic fit
\begin{equation}\beta \mu_{\cal E} \approx \frac{1}{2}+
\frac{\epsilon(1)}{k_B T}\label{hyfit}
\end{equation} is the dash-dotted line in Fig. \ref{fig:hyp}. At high-temperature, the energy/temperature term reaches the
horizontal asymptote of the hyperbole, that, as predicted by the
high temperature approximation, is equal to $0.5$. However, in the
transition region, say $0.3 < T <3$, the approximation error is
appreciable. For instance, for $T=1$ the numerical calculation
gives $\beta \mu_{\cal E}=2.37$, while the heuristic hyperbolic
fit (\ref{hyfit}) gives $\beta \mu_{\cal E}=2.86$.

Figs. \ref{fig:piab30} and \ref{fig:corr30} report the results
obtained numerically for $N=30$ three-dimensional particles. The
number of energy levels considered in the triple sum of equation
(\ref{3dexact}) is 15, which generates $15^3$ terms of the triple
sum, but we have verified that also only 5 terms do virtually not
impact graphs. Fig. \ref{fig:piab30} shows that the entropy of
indexed particles and, consequently, the entropy of non indexed
particles, which must be between zero and the entropy of indexed
particles, drops virtually to zero at the ''cutoff'' temperature
of about 1.3 Kelvin, where the entropy of the 30 three-dimensional
indexed particles is 1.44 (1.44/90 per degree of freedom), while
the entropy of the non-indexed particles is 1.18. A so low value
of entropy is compatible with the fact, already pointed out, that
the expected energy per particle is close to that of the ground
state, meaning that particles are with high probability in the
ground state. Fig. \ref{fig:corr30} reports the relative quantum
correction, that is
\[\frac{\sum_{\bar{l}^3
\in \{\bar{\cal L}^3\}}\braket{\log({\cal
N}_{\Sigma}(\bar{l}^3)!)}}{\log(N!)},\] versus temperature. The
graph shows that, for temperature below 1 Kelvin, the numerator of
the above relative quantum correction is virtually equal to the
denominator, hence that the third term of (\ref{3dexact})
completely compensates the term $-\log(N!)$, preventing negative
values of the entropy. As expected, the quantum correction tends
quickly to zero at high temperature. At $10$ Kelvin it is still
appreciably above zero, but at $100$ Kelvin it becomes negligible.

\section{Conclusions}

A step towards a full information-theoretic paradigm for the
statistical mechanics of systems made by independent particles has
been proposed in the paper. This approach, we would say, this
vision, is based on the identification of information theoretic
concepts as typicality, conditional entropy, mutual information,
Holevo accessible information, with the corresponding physically
equivalent quantities. Our paradigm rules out the postulate of
equally probable microstate, which has been object of debate since
when it was originally proposed by Einstein. The renounce to the
postulate not only is compatible with all the results that we know
about many-particle systems, but also leaves space to an harmonic
and unified treatment of low and high temperature, small and large
number of particles, systems of indexed and of non-indexed
particles.

Much work is to be done along the direction taken by this paper.
We hereafter mention few points that are not covered in this paper
and that could be considered in the future. One point that needs
investigation is the role of the assumption of identical
distribution. For systems at the thermal equilibrium, it seems
that this assumption is reasonable, but, even if in principle this
assumption could be made also for a system that is not at the
thermal equilibrium, it could become questionable and could lead
to misleading conclusions when the system evolves towards
equilibrium through to non-reversible transformations. For
instance, when two gases at different temperatures are mixed, can
de mix be described by one probability distribution obtained by
the weighted sum of the two and, if so, till what extent? Finally,
we mention that another open point is how and if the Bose-Einstein
condensed state can be inserted inside the framework sketched in
this paper.

\section{Appendix A: Random number of particles}

Let us consider the case where the system can randomly exchange
particles with the environment. This happens, for instance, when
the system occupies a region of space that is not bounded by
impenetrable walls, so that particles can freely come in and out
of the region of interest. Let us start the analysis by observing
that, in the case of a fixed number of particles, the probability
distribution of the random state can be written by multiplying the
multinomial probability distribution by a delta-type distribution
that imposes the total number of particles:
\begin{align} & p_{\bar{\cal N}_{\Sigma}}(\bar{n}_{\Sigma})=
\delta\left(N-\sum_{l \in \{{\cal L}\}}
{n}_{\Sigma}(l)\right)N!\prod_{l \in \{{\cal L}\}}\frac{(p_{\cal
L}(l))^{{n}_{\Sigma}(l)}}{{n}_{\Sigma}(l)!}.\nonumber\end{align}
If, in place of being deterministic, the total number of particles
is random, then probability distribution $p_{{\cal N}_{\Sigma}}$
of the total random number of particles takes the place of the
delta-type distribution and the factorial of the fixed number of
particles becomes the factorial of the random number of particles,
leading to
\begin{align} & p_{\bar{\cal N}_{\Sigma}}(\bar{n}_{\Sigma})=
p_{{\cal N}_{\Sigma}}(n_{\Sigma}){n_{\Sigma}}!\prod_{l \in \{{\cal
L}\}}\frac{(p_{\cal
L}(l))^{{n}_{\Sigma}(l)}}{{n}_{\Sigma}(l)!}.\label{pd}\end{align}
We assume that the position where a particle can be detected is a
continuous random variable with uniform distribution over a
virtually unbounded region of space that includes the region of
interest and the surroundings. This assumption is reasonable for
systems at the thermal equilibrium, in which case the system is
one element of the grand canonical ensemble. If, besides the
uniform distribution, we also assume that two particles cannot be
detected in the same position and that the positions where
particles are detected are independent random variables, then the
random number ${\cal N}_{\Sigma}(l)$ of particles that occupy the
$l$-th energy level inside the region of interest follows the
Poisson distribution, therefore
\[p_{{\cal N}_{\Sigma}(l)}(n_{\Sigma}(l))=
\frac{(\lambda(l))^{{n}_{\Sigma}(l)}e^{-\lambda({l})}}
{{n}_{\Sigma}(l)!},\] where $\lambda(l)$ is the expected number of
particles inside the region of interest that occupy the $l$-th
energy level. Since the sum of Poisson random variables is a
Poisson random variable,
\[p_{{\cal N}_{\Sigma}}(n_{\Sigma})=
\frac{(\lambda)^{{n}_{\Sigma}}e^{-\lambda}}{{n}_{\Sigma}!},\]
where
\[\lambda=\sum_{l \in \{{\cal L}\}}\lambda(l).\]
The Poisson distribution has been used in the past in very many
cases. A case that is worth mentioning because it looks very close
to ours is that of London's bombing in the second war, where the
random number of bombs (in \cite{londonbombing} the distributions
of V-1 bombs and of V-2 rockets, the ''flying bombs,'' are
analyzed) falling in a sub-area of an area subject to almost
uniform bombing is modelled by the Poisson distribution. Here, in
place of the random number of bombs of the two types we have the
random number of particles of different energies. It is easy to
see that the entries of the random vector $\bar{\cal N}_{\Sigma}$
are independent between them. Independency between the occupancy
numbers is not that obvious. For instance, the occupancy numbers
of the multinomial distribution are not independent between them,
because their sum is forced to be equal to a deterministic number.
Substituting the Poisson distribution for $p_{{\cal
N}_{\Sigma}}(n_{\Sigma})$ in (\ref{pd}) one gets
\begin{align} & p_{\bar{\cal N}_{\Sigma}}(\bar{n}_{\Sigma})=
\lambda^{n_{\Sigma}}e^{-\lambda}\prod_{l \in \{{\cal
L}\}}\frac{(p_{\cal
L}(l))^{{n}_{\Sigma}(l)}}{{n}_{\Sigma}(l)!}.\label{pd2}\end{align}
Substituting
\[\lambda^{n_{\Sigma}}=\prod_{l \in \{{\cal L}\}}
\lambda^{{n}_{\Sigma}(l)}(l) ,\]
\[e^{-\lambda}=\prod_{l \in \{{\cal L}\}}e^{-\lambda({l})}\] and
\begin{equation}\lambda(l)=p_{\cal L}(l) \lambda\label{plam}\end{equation} in equation (\ref{pd2}), one
finds
\begin{align} & p_{\bar{\cal N}_{\Sigma}}(\bar{n}_{\Sigma})=
\prod_{l \in \{{\cal
L}\}}\frac{(\lambda(l))^{{n}_{\Sigma}(l)}e^{-\lambda(l)}}{{n}_{\Sigma}(l)!}.\label{pd3}\end{align}
The probability distribution (\ref{pd3}) is the product of Poisson
distributions with parameters $\lambda(l)$, hence the individual
random variables are independent between them. The random entropy
and the entropy of the Poisson random variable ${\cal N}$ with
parameter $\lambda$ are
\begin{equation}H({\cal
N})=-{\cal N}\log(\lambda)+\lambda+\log({\cal N}!),\nonumber
\end{equation}
\begin{equation}H_{\cal
N}=-\lambda(\log(\lambda)-1)+\braket{\log({\cal N}!)},\nonumber
\end{equation} where
\begin{align}  \braket{\log({\cal N}!)}
&=e^{-\lambda}\sum_{n=0}^{\infty}\frac{\lambda^n \log(n!)}{n!}.
\nonumber
\end{align}
An integral form for the above expectation has been recently
worked out in \cite{mahdi}. The Shannon entropy of the vector of
independent random variables is the sum of the individual
entropies, therefore
\begin{align} H_{\bar{\cal N}_{\Sigma}}
&=\sum_{l \in \{{\cal L}\}} H_{{\cal N}_{\Sigma}(l)} \nonumber \\&
=\lambda H_{{\cal L}}-\lambda(\log(\lambda)-1) +\sum_{l \in
\{{\cal L}\}}\braket{\log({\cal N}_{\Sigma}(l)!)} \nonumber \\ &
\approx \lambda H_{{\cal L}}-\log(\lambda!) +\sum_{l \in \{{\cal
L}\}}\braket{\log({\cal N}_{\Sigma}(l)!)}, \nonumber
\end{align}
where in the second equality we use (\ref{plam}) and, in the last
line, which is obtained by Stirling's formula, for non-integer
$\lambda$ one uses the gamma function in place of the factorial.
Putting $\lambda=N$ we immediately appreciate the strong
similarity between the above entropy and that of the multinomial
random variable (\ref{ei}). The conclusion that, when the system
exchanges particles with the environment, the probability
distribution of ${\cal N}_{\Sigma}(l)$ is Poisson, is achieved, by
different arguments, also in the standard treatment of the
classical ideal gas at high temperature, see for instance
21.2.4-21.2.6 of \cite{sek} where, however, independency between
the random occupancy numbers is not discussed. Here the Poisson
distribution for the random number of particles descends directly
from the uniformity of the spatial distribution of particles in
the unbounded region of space that includes the region of
interest, an assumption that seems reasonable also at low
temperature.

\section{Appendix B: randomness of macrostates}

Deterministic constraints imposed on the system must be
distinguished from macrostates. A deterministic constraint is the
expectation of the corresponding macrostate divided by the number
of particles. The most detailed deterministic constraint is the
entire probability distribution,
\begin{align}p_{\cal L}(l)&= \frac{1}{N}\braket{{\cal N}_{\Sigma}(l)}, \
\forall \ l \in \{{\cal L}\}.\nonumber \end{align} The probability
distribution allows for the computation of any other deterministic
constraint, exactly as the occupancy numbers do with any other
random macrostate through (\ref{randomex}). Our distinction
between deterministic constraint and macrostate is compatible with
the Gibbsian view of a {\em canonical ensemble} of systems, where
the systems of the ensemble are assumed to be at the thermal
equilibrium at a given temperature and the number of particles is
the same for all the systems of the ensemble, therefore the
deterministic constraint is the Boltzmann distribution with a
prescribed number of particles and a prescribed expected energy
per particle that depends on the temperature, while the random
macrostate is system's total energy which, in the random
experiment of picking the system of interest from the ensemble,
comes together with the picked system. It should be added that,
with the assumption of i.i.d. particles, we don't need an ensemble
of systems: an ensemble of particles is all we need if we build
the system of interest by randomly picking $N$ particles from the
ensemble of particles. This ensemble of particles is equivalent to
the Gibbsian canonical ensemble of systems and the two are
compatible with the time evolution of an ergodic system resulting
from the random interaction of the system with the surrounding
environment.

Randomness of macrostates in the canonical ensemble is universally
recognized since the times of Maxwell. The randomness that is
present in the total system energy makes it possible that system's
energy fluctuates around its expectation. One can figure out the
extreme situation where a system made by only one particle is in
contact with the heath bath. Even if, occasionally, the energy of
the one-particle system is below (above) what expected, it can get
even lower (higher) if, in the ''contact'' with the heath bath,
the particle of the system kicks a slower (faster) particle of the
heath bath. Quoting from \cite{myrvold}: {\em As Maxwell points
out, measurable thermodynamic quantities are averages over many
molecular quantities; if the molecular quantities exhibit
fluctuations that are probabilistically independent of each other,
these fluctuations will tend to be washed out as the number of
molecules considered is increased.}

However, the consequences of this randomness seem not to be fully
exploited in the standard treatment of statistical mechanics, at
least in the sense pursued in this paper. Specifically, we cannot
find equation (\ref{zeroentropy57}), that for systems of the
canonical ensemble reads
\begin{equation}
H_{\bar{\cal L}}=H_{\bar{\cal L}|{\cal E}_{\Sigma}}+H_{{\cal
E}_{\Sigma}},\label{zupa} \end{equation} in anyone of the many
textbooks that we consulted, where, in fact, also the conditional
probability distribution $p_{\bar{\cal L}|{\cal E}_{\Sigma}}$ is
overlooked. All these textbooks deeply consider the canonical
entropy $H_{\bar{\cal L}}$ and the microcanonical entropy
$H_{\bar{\cal L}|{\epsilon}_{\Sigma}}$. The canonical system
imports macrostate's randomness through the Boltzmann
distribution, an approach that does not need to explicitly show
macrostate's probability distribution and macrostate's entropy. In
the microcanonical system, the ''given'' macrostate
${\epsilon}_{\Sigma}$ is treated as a deterministic condition that
is often taken equal to the expectation of system's energy. In the
end, neither the canonical approach nor the microcanonical
approach explicitly analyze macrostate's probability distribution
and, with it, macrostate's entropy. This has made impossible in
the past, with the exception of \cite{zupa} that we are going to
discuss, to explain with a simple formula like (\ref{zupa}) the
relation between the canonical entropy and the microcanonical
entropy and, more generally, between the Gibbs entropy and the
Boltzmann-Planck entropy.

We argue that the incomplete consideration of macrostates'
randomness originates from a fallacy in the argument that supports
the last part of the passage of \cite{myrvold} quoted above. In
thermodynamics, where $N$ is large, the random sum that
characterizes the random macrostate becomes virtually equal to its
expectation within a small {\em relative} random error, which, in
the case of the random total entropy, can be seen as the random
version of the deterministic $\eta$ that appears in
(\ref{conv2})-(\ref{narrow3}). At the same time, for $N
\rightarrow \infty$, equation (\ref{lack}) shows that macrostate's
entropy {\em per particle} becomes vanishingly small. At a first
glance, this argument could seem convincing because, physically
speaking, one could think that the relative error and the entropy
per particle are what really matters to all practical purposes.
However, it remains that, for $N \rightarrow \infty$, the random
sum does not converge to any deterministic value. On the contrary,
if we represent the randomness of the random sum with its standard
deviation, we see that it increases with the square root of the
number of particles. Therefore the fallacy of this argument is
that it is not true that this randomness is washed out when $N
\rightarrow \infty$, what is washed out is the {\em relative}
randomness. The distinction between absolute and relative
randomness is not a mere sophism or a subtle technicality that has
no impact on reality. First, the contribution of the term
$H_{{\cal E}_{\Sigma}}$ to system's entropy $H_{\bar{\cal L}}$ can
significantly impact the entropy of systems of few particles, till
to become the only contribution to the entropy of a system made by
one particle, for which macrostate and microstate coincide.
Second, in a forthcoming section of this paper we will show that
randomness of macrostates is the only randomness that contributes
to the macroscopic entropy of systems of a large number of
non-indexed particles.

Paper \cite{zupa} is the only one, among the very many papers that
we studied during this research, that fully catches randomness of
macrostates in the sense discussed above. Specifically, what the
authors of \cite{zupa} call {\em the probability distribution of
the macroscopic state of the system} and denote $P(a)$ can be
identified with our $p_{\bar{\cal N}_{\Sigma}}(\bar{n}_{\Sigma})$
and what they call $W(a)$ can be identified with our
$W_M(\bar{n}_{\Sigma})$. With this identification, our
$H(\bar{\cal L}|\bar{n}_{\Sigma})$ is what they call {\em
Boltzmann entropy} in their eqn. (6) and our $H_{\bar{\cal
L}|\bar{\cal N}_{\Sigma}}$ is what they call {\em the mean value
of the Boltzmann entropy} in their eqn. (7). Then they use this
mean value of Boltzmann entropy in the entropic relation
(\ref{zeroentropy57}). In our opinion, paper \cite{zupa} is
particularly important because, to our best knowledge, it is the
first paper that makes use in a physical context of the concepts
of random entropy and of conditional entropy and that recognizes
that Gibbs entropy $H_{\bar{\cal L}}$ is the sum of the term
$H_{\bar{\cal N}_{\Sigma}}$, which is due to the randomness of the
macrostate, and of the term $H_{\bar{\cal L}|\bar{\cal
N}_{\Sigma}}$, which is due to the randomness of the microstates
whose union form the macrostate. However, it should be added that
the authors of \cite{zupa} seem not to be aware that their
Boltzmann entropy is a conditional entropy. As a matter of fact,
they do not make any explicit use of conditional probability
distributions. They hide the conditional probability distribution
$p_{\bar{\cal L}|\bar{\cal N}_{\Sigma}}$ behind the corresponding
multinomial coefficient $W_M(\bar{n}_{\Sigma})$ and implicitly
consider understood the equality (\ref{multcond}). Also, they
claim but don't prove that the probability distribution of the
occupancy numbers is multinomial, neither they use the multinomial
distribution, in fact they consider the Gaussian approximation in
place of the exact multinomial distribution. Finally, the big
difference with our paper is that the authors of \cite{zupa} still
{\em assume} that microstates are conditionally equiprobable.

\section{Appendix C: information-theoretic typicality}
The commonly accepted definition of entropy in statistical
mechanics and thermal physics is
\begin{equation}
S=k_B \log(W),\label{bp}
\end{equation}
where $S$ is the entropy, $\log(x)$ is the natural logarithm of
$x$, \[k_B=1.38  \cdot 10^{-23} \ \ \mbox{J} \cdot \mbox{K}^{-1}\]
is Boltzmann's constant and $W$ is the number of microstates that
are compatible with the constraints imposed on the system. The
phrase of Von Neumann that opens the introduction of this paper
reflects the fact that the above entropy, called by many authors
Boltzmann-Planck entropy, has been for a long time and is still
today object of discussion. Actually, the right side of the
equality is an entropy only if the $W$ microstates are equally
probable but, still today, there is not any mathematical proof of
microstates' equiprobability. Nonetheless, equation (\ref{bp}) is
universally recognized to be the very foundation of thermal
physics and statistical mechanics, that, in the absence of a
mathematical proof but, at the same time, in the presence of
strong physical evidences, take it as a postulate. Einstein was
the first who, in \cite{einstein1909}, suggested to postulate or,
using the original German verb, {\em verlangt}, equiprobability of
microstates: {\em Neither Mr. Boltzmann nor Mr. Planck gave a
definition of W. They put purely formally W = number of
complexions of the state under consideration. If one now demands
that these complexions be equally probable...}

There is wide skepticism in the research community about the
possibility of producing a mathematical proof of microstates
equiprobability. This skepticism dates back to the following
quotation from Khinchin \cite{Khi}, that is widely reported in the
open bibliography, e.g. \cite{campisi,wright}: {\em All existing
attempts to give a general proof of this postulate must be
considered as an aggregate of logical and mathematical errors
superimposed on a general confusion in the definition of the basic
quantities.} About fifteen years ago, by arguments based on
typicality, the two papers \cite{typpopescu} and \cite{typgold}
proved that microstates cannot be equiprobable, but, at the same
time, they showed that microstates converge to equiprobability as
the number of particles grows. Paper \cite{typpopescu} claims in
the abstract that {\em ...the main postulate of statistical
mechanics, the equal a priori probability postulate, should be
abandoned as misleading and unnecessary.} This claim is
substantiated by a theorem based on the law of large numbers. The
two papers \cite{typpopescu} and \cite{typgold} have attracted
noticeable interest, see for instance \cite{typbart}, and can be
considered a cornerstone of modern quantum thermodynamics, see
\cite{quantumthermo,trends}. A result similar to that of
\cite{typpopescu} and \cite{typgold} has been recently published
by this author in \cite{spalvieri}. The approach of
\cite{spalvieri}, although relying on typicality as
\cite{typpopescu} and \cite{typgold}, is much more
information-theoretically oriented than \cite{typpopescu} and
\cite{typgold}. Hereafter we report the main points of
\cite{spalvieri}.

 When the $N$ random variables of the random vector
$\bar{\cal L}^N$ are i.i.d. we have
\begin{align}\lim_{N \rightarrow \infty } \frac{1}{N}
H(\bar{\cal L}^N)&=
 \lim_{N \rightarrow \infty }- \frac{1}{N}\log(p_{\bar{\cal L}^N}
 (\bar{\cal L}^N))
\nonumber \\ &= \lim_{N \rightarrow \infty }-
\frac{1}{N}\sum_{i=1}^N\log(p_{{\cal L}_i}({\cal
L}_i))\label{inda}
\\ &= \lim_{N \rightarrow \infty }-
\frac{1}{N}\sum_{i=1}^N\log(p_{{\cal L}}({\cal L}_i))\label{iddi}
 \\ & =  H_{{\cal L}},\label{syse}\end{align} where (\ref{inda}) is the
assumption of independency, (\ref{iddi}) is the assumption of
identical distribution, and (\ref{syse}) is the law of large
numbers. The convergence of the limit (\ref{inda}) to (\ref{syse})
is in various senses. Our main concern is the case of large but
finite $N$, hence we consider convergence in probability, that is
\begin{equation}\hspace{-0.0cm} \lim_{N \rightarrow \infty}
Pr\left(\left|\frac{1}{N} \log(p_{\bar{\cal L}^N}(\bar{\cal
L}^N))+H_{{\cal L}} \right|> \eta \right) = 0, \label{conv2}
\end{equation} for every $\eta
>0$. The above limit is a pillar of classical information theory.
In chapter 3 of their book \cite{cover}, Cover and Thomas call it
the {\em Asymptotic Equipartition Property} (AEP).\footnote{Here
it is the probability that is equally partitioned, not the energy,
hence this AEP has nothing to do with the classical energy
equipartition property.} In the context of ergodic theory, many
authors refer to (\ref{conv2}) as to the Shannon-McMillan theorem.
Almost everywhere convergence, that strengthens convergence in
probability, is known as the Shannon-McMillan-Breiman theorem.

Starting from (\ref{conv2}) we arrive at the concept of {\em
typical set} $\{\bar{\cal T}_{\eta}^N\}$, which is a subset of the
set $\{\bar{\cal L}^N\}$,
\begin{equation}
\{\bar{\cal T}_{\eta}^N\} \subseteq \{\bar{\cal
L}^N\},\label{subset}
\end{equation}
whose definition is based on the random probability (\ref{ranp}).
Specifically, the typical set is made by those random vectors
whose random probability, for every $\eta$ and for sufficiently
large $N$, is a random number that lies in the narrow range
\begin{equation}e^{-N(H_{{\cal L}}+\eta)} \leq p_{\bar{\cal
L}^N}(\bar{\cal T}^N_{\eta}) \leq e^{-N(H_{{\cal
L}}-\eta)}.\label{narrow}
\end{equation}
The properties of the typical set are that the number of its
elements is in the narrow range
\begin{equation}e^{N(H_{{\cal L}}+\eta)} \geq |\bar{\cal T}^N_{\eta}|
\geq (1-\eta)e^{N(H_{{\cal L}}-\eta)}\label{narrow2}
\end{equation}
and that the probability that the outcome of the random vector
$\bar{\cal L}^N$ belongs to the typical set is
\begin{equation}Pr(\bar{\cal L}^N \in \{\bar{\cal T}^N_{\eta}\})>
1-\eta, \label{narrow3}
\end{equation}
see again \cite{cover}. Using the language of statistical
mechanics, we can look at the elements of the typical set as at
the {\em accessible} microstates. For $N$ large enough,
inequalities (\ref{narrow})-(\ref{narrow3}) show that their number
is nearly $e^{N H_{{\cal L}}}$ and that their {\em overwhelming
majority} is nearly equiprobable. Compared to the standard
approach, where microstates are either accessible (with uniform
probability) or non-accessible (zero probability of access), here
all the microstates that have non-zero probability are in
principle accessible, but with different probabilities.

\section{Appendix D: Historical positioning of the paper}

\subsection*{Ideal gas in a container}

Looking at (\ref{ei}), we see that the term $-\log(N!)$ is the one
introduced by Gibbs to make the phase-space entropy of non-indexed
particles compatible with his famous paradox.  The genesis of the
term $-\log(N!)$ has been in the past object of debate. Many
authors report that this term was introduced by Gibbs as an {\em
ad hoc} correction of the entropy of indexed particles to ensure
extensivity of entropy, but Jaynes argues in \cite{jgibbs} that an
early work of Gibbs contains an analysis that correctly explains
the presence of the term $-\log(N!)$ in the entropy formula.

After Gibbs, Tetrode followed a reasoning that brought him very
close to (\ref{ei}). He considered quantization of the phase space
into cells of area $2 \pi \hbar$  and argued that gas states
obtained by permutation of particles should not be counted as
different, see \cite{grimus}. At high temperature all the
particles occupy different cells of the phase space, hence the
number of permutations expressed by the multinomial coefficient
becomes equal to $N!$, whose logarithm is just the correction term
considered by Tetrode. Basically, equation (\ref{ei}) can be seen
as a consequence of Tetrode's approach, it is enough to consider
the random multinomial coefficient as it is and then to take the
expectation of its logarithm. The denominator of the multinomial
coefficient, overlooked by Tetrode, generates the sum that appears
in (\ref{ei1}). This sum becomes important at low temperature,
where many particles occupy the same quantum state, reducing
significantly below $N!$ the number of distinct permutations. If
this term is overlooked, at temperature low enough the correction
represented by $-\log(N!)$ becomes too large and the entropy falls
below zero, as it actually happens in the Sackur-Tetrode entropy
formula. The difference with our approach is that in the
Sackur-Tetrode approach what is quantized is the phase space,
while here we consider a truly quantum approach and, in place of
defining entropy as a function of position and momentum, we define
it as a function of the discrete classical random vector
$\bar{\cal L}$. Compared to the phase-space approach, which needs
approximations to take into account quantum corrections at low
temperature, see the recent paper \cite{psapprox}, the
consideration of a quantum system natively allows to exactly
import the quantum corrections inside the model, leading to an
exact description of entropy also at low temperature.

Einstein showed in \cite{einstein24} that the term $-\log(N!)$ is
inherently present in the entropy of quantum gases, without the
need of any {\em ad hoc} intervention. After Einstein, it became
standard to treat the classical ideal gas at high temperature as a
quantum gas, thus incorporating the term $-\log(N!)$ in the
entropy. Actually, at high temperature the effect of quantization
is negligible therefore there is virtually no distinction between
classical and quantum gases. However, both the Sackur-Tetrode
approach and the full-quantum approach describe the ideal gas only
at high temperature, while (\ref{ei}) gives the entropy of the
ideal gas at low and high temperature.

Swedensen in \cite{swe} defines the entropy in the quantized phase
space by embedding in the definition the term $-\log(N!)$. The
crucial point of Swedensen's definition is that the probability
distribution of finding a random number of particles in
sub-regions of space is multinomial (even if Swedensen's example
is based on the binomial distribution, the extension to the
multinomial distribution is straightforward). The term $-\log(N!)$
is then imported inside the entropy just from the multinomial
distribution, exactly as it happens in our case.

\subsection*{Energy macrostate of a system of monochromatic quantum harmonic oscillators}
The random total quantum number ${\cal L}_{\Sigma}$,
\begin{equation}{\cal L}_{\Sigma}=\sum_{i=1}^{N}{\cal L}_{i},
\nonumber \end{equation} is the sum of $N$ i.i.d. geometric random
variables, therefore its distribution is the negative binomial
distribution:
\begin{align}
p_{{\cal L}_{\Sigma}}(l_{\Sigma})&=W_{NB}({l}_{\Sigma})
p^N(1-p)^{{l}_{\Sigma}},\nonumber
\end{align}
where ${l}_{\Sigma}$ spans the support set of ${\cal L}_{\Sigma}$
and \begin{align}   W_{NB}({l}_{\Sigma})&=
\frac{({l}_{\Sigma}+N-1)!}{{l}_{\Sigma}!(N-1)!}
 \label{who}\end{align}
is the negative binomial coefficient, which Planck calls $R$ after
equation (4) of \cite{planck}. The interested reader is referred
to chapter 5 of \cite{univariate} for the relation between the
geometric distribution and the negative binomial distribution and
to \cite{mahdi} for the entropy of the negative binomial
distribution. To our best knowledge, it is recognized in
\cite{mandel} for the first time that the probability distribution
of the random total quantum number ${\cal L}_{\Sigma}$ of a system
of $N$ i.i.d. quantum harmonic oscillators all oscillating with
the same angular frequency is negative binomial. Equation
(\ref{wsumenergy}) becomes
\begin{equation}W_{NB}(l_{\Sigma})=\sum_{\bar{n}_{\Sigma}: \sum_{l
\in \{{\cal L}\}}l{n}_{\Sigma}(l)={l}_{\Sigma}
}W_M(\bar{n}_{\Sigma}).\label{wsum}
\end{equation} Boltzmann already pointed out that the negative
binomial coefficient, that he calls $J$ in \cite{boltz}, is {\em
the sum of all possible complexions}.

As the multinomial coefficient, also the negative binomial
coefficient is a recurring figure in statistical mechanics and
quantum mechanics. It can be calculated by the {\em stars and
bars} method, which, as reported in \cite{monaldi}, was introduced
in 1914 in a paper by Ehrenfest and Kamerlingh-Onnes and was later
brought by Feller \cite{feller} in a context different to physics.
The stars and bars method counts the number of distinct
arrangements of ${l}_{\Sigma}$ stars (energy quanta) into $N$
cells (oscillators) separated by $N-1$ bars plus two additional
bars at the beginning and at the end of the pattern. This number
is the number of distinct vectors $\bar{l}^N$ whose sum is
${l}_{\Sigma}$.

Thanks to the negative binomial distribution, Mandel {\em
demonstrates} in \cite{mandel} that all the $W_{NB}({\cal
L}_{\Sigma})$ arrangements of energy quanta are {\em conditionally
equiprobable} given the total number ${\cal L}_{\Sigma}$ of energy
quanta. Here the random condition is ${\cal L}_{\Sigma}$ and, with
arguments analogous to those that led to (\ref{deltaw}), for the
conditional probability of microstates we get
\begin{align}  p_{\bar{\cal L}|{\cal L}_{\Sigma}}(\bar{l},l_{\Sigma})=\left\{
\begin{array}{cc} \frac{1}{W_{NB}(l_{\Sigma})}
, &  \bar{l} \in \{\bar{l}(l_{\Sigma})\}, \\
0,  &  elsewhere,\\
\end{array} \right.
\label{pcondwho}\end{align} This conditional equiprobability has
the same interpretation as (\ref{deltaw}). What changes between
the two is only the random condition and its probability
distribution. Since
\begin{align} H({\cal L}_{\Sigma}|\bar{\cal L})=
0, \nonumber
\end{align}
\begin{align} H_{{\cal L}_{\Sigma}|\bar{\cal L}}=
0, \nonumber
\end{align}
the entropic relation analogous to  (\ref{zeroentropy57}) is
\begin{align} H_{\bar{\cal L}}-H_{
\bar{\cal L}|{\cal L}_{\Sigma}}= H_{{\cal L}_{\Sigma}} \geq
0,\nonumber\end{align} where the inequality is satisfied with
equality only when ${\cal L}_{\Sigma}$ is actually deterministic,
that is, at the ground state, where ${\cal L}_{\Sigma}=0$.

\subsection*{Boltzmann's total number of complexions and
Planck's entropy}

Planck takes Boltzmann's total number of complexions (\ref{who}),
forcing inside it the mean value of the total quantum number
\begin{equation}\braket{{\cal L}_{\Sigma}}= N \mu_{\cal L},\label{pldet} \end{equation} see \cite{planck}, see also
16.3 of \cite{sek} and 3.8 of \cite{pathria}. The gamma function
can be used in the factorials in case of non-integer
$\braket{{\cal L}_{\Sigma}}$. Then Planck, citing Boltzmann,
claims that the logarithm of the above $W_{NB}(\braket{{\cal
L}_{\Sigma}})$ is entropy, even if Boltzmann never assigned to the
logarithm of the above $W_{NB}(\braket{{\cal L}_{\Sigma}})$ the
role of entropy. In fact, nobody understands why the logarithm of
the negative binomial coefficient should be entropy. But,
undoubtedly, it works.

A question naturally arises. Bose's entropy\footnote{We consider
here only one frequency, so Boses's index $s$ here disappears. The
rest is mapped as follows, Bose on the left, this paper on the
right of the arrow.
\begin{itemize}
\item  $A$ cells in phase space $ \rightarrow$ $N$ oscillators,
 \item number of cells at the $l$-th
energy level, $p_l \rightarrow   N p_{\cal L}(l)$, \item $W
\rightarrow W_M(\braket{\bar{\cal N}_{\Sigma}})$,
 \item $k_B$ times the temperature, $\beta
\rightarrow \beta^{-1},$ \item energy of one quantum, $h \nu
\rightarrow \hbar \omega,$ \item expected total number of energy
quanta, $N \rightarrow N\mu_{\cal L}=\mu_{{\cal L}_{\Sigma}},$
\item expected total energy, $E \rightarrow N\mu_{\cal E}$, \item
$B \rightarrow Np,$ this is only a service variable.
\end{itemize}
The entropy, calculated by Bose as $\log(W_M(\braket{\bar{\cal
N}_{\Sigma}}))$, appears in the second-top equation of the last
column of \cite{bose}. In passing, we point out that the sign of
the argument of the exponential in Bose's second-last equation
that expresses the entropy is flipped. Finally, note that Bose
puts to zero the energy of the ground state, while here we put it
to $\omega \hbar /2$.} \cite{bose} $\log(W_M(\braket{\bar{\cal
N}_{\Sigma}}))$ and Planck's entropy $\log(W_{NB}(\braket{{\cal
L}_{\Sigma}}))$ are the logarithms of two different numbers, how
can they be the same entity? Luckily, as Von Neumann said to
Shannon, no one really knows what entropy really is, therefore
everyone can feel free to put inside that logarithm whatever he
likes. Provided that it works, we add. In fact, for large $N$, the
two logarithms are both a good approximation to the Gibbs entropy,
because in (\ref{wsum}) the term $W_M(\braket{\bar{\cal
N}_{\Sigma}})$ dominates the sum, see problem 3.4 of
\cite{pathria}. What changes is the accuracy of the approximation,
that, for finite $N$, can be different in the two cases.
Specifically, Planck's entropy can be manipulated as follows
\begin{align}
& \log (W_{NB}(\braket{{\cal L}_{\Sigma}})) \approx   \log
\left(\frac{(\braket{{\cal L}_{\Sigma}}+N)!}{\braket{{\cal
L}_{\Sigma}}!N!}\right) \nonumber
\\ &
\stackrel{\text{Stirling}}{\approx}  \braket{{\cal
L}_{\Sigma}}\log\left(1+\frac{N}{\braket{{\cal L}_{\Sigma}}}
\right)+N \log\left(1+\frac{\braket{{\cal L}_{\Sigma}}}{N} \right)
\nonumber
\\ & =  N (\mu_{\cal L}\log(1+\mu_{\cal L}^{-1})+
\log(1+\mu_{\cal L})) \nonumber \\
&= H_{\bar{\cal L}}\nonumber.
\end{align}
Compared to (\ref{shmulti}), the above equation shows why Planck
can use the negative binomial coefficient in place of the
multinomial coefficient.

\end{document}